\documentclass[twocolumn,floatfix,prx,aps,showpacs]{revtex4-2}
\usepackage{graphicx,amsmath,amssymb,color,nicefrac,multirow, makecell, ragged2e,float}
\usepackage[unicode=true,colorlinks=true,linkcolor=blue,citecolor=blue,urlcolor=blue]{hyperref}
\usepackage[titletoc,title]{appendix}

\newcommand{\be}{\begin{equation}}
\newcommand{\ee}{\end{equation}}

\newcommand{\ba}{\begin{eqnarray}}
\newcommand{\ea}{\end{eqnarray}}

\def\beq{\begin{eqnarray}}
\def\eeq{\end{eqnarray}}

\makeatletter
\newcommand*{\rom}[1]{\expandafter\@slowromancap\romannumeral #1@}
\makeatother

\newcommand{\non}{\nonumber\\}
\usepackage{bm,physics,enumerate,booktabs,xcolor,float}

\allowdisplaybreaks[4]

\begin{document}
\title{Exactly Solvable Hamiltonian for Non-Abelian Quasiparticles}
\author{Koji Kudo$^1$, A. Sharma$^1$, G. J. Sreejith$^2$ and J.K. Jain$^1$}
\affiliation{$^1$The Pennsylvania State University, 104 Davey Lab, University Park, Pennsylvania 16802, USA}
\affiliation{$^2$Indian Institute of Science Education and Research, Pune 411008, India}
\begin{abstract}
Particles obeying non-Abelian braid statistics have been predicted to emerge in the fractional quantum Hall effect.
In particular, a model Hamiltonian with short-range three-body interaction ($\hat{V}^\text{Pf}_3$) between electrons confined to the lowest Landau level provides exact solutions for quasiholes, and thereby allows a proof of principle for the existence of quasiholes obeying non-Abelian braid statistics. 
We construct, in terms of two- and three- body Haldane pseudopotentials, a model Hamiltonian that can be solved exactly for both quasiholes and quasiparticles, and provide evidence of non-Abelian statistics for the latter as well. The structure of the quasiparticle states of this model is in agreement with that predicted by the bipartite composite-fermion model of quasiparticles with exact lowest Landau level projection.
We further demonstrate adiabatic continuity for the ground state, the ordinary neutral excitation, and the topological exciton as we deform our model Hamiltonian continuously into the lowest Landau-level $\hat{V}^\text{Pf}_3$ Hamiltonian.  
\end{abstract}

\maketitle

\section{Introduction}

The phenomenon of the fractional quantum Hall effect (FQHE)~\cite{Tsui82}, which occurs when electrons confined to two-dimensions are subjected to a strong magnetic field, has proved to be a treasure-trove of exotic emergent structures. According to the Moore-Read (MR) proposal~\cite{Moore91,Read00}, the FQHE at filling factor $\nu= 5/2$~\cite{Willett87} arises from a succession of remarkable emergences: First, electrons in the first excited Landau level (LL), which is half full, bind two vortices each to form composite fermions (CFs)~\cite{Jain89,Jain07}. CFs experience no net magnetic field and attempt to form the Halperin-Lee-Read Fermi sea, as they are known to do in the lowest LL (LLL)~\cite{Halperin93}. The Fermi sea in the first excited LL, however, is unstable to a topological p-wave pairing of CFs, which opens a gap to produce FQHE. Furthermore, the Abrikosov vortices of this paired state are thought to harbor Majorana zero modes, which obey non-Abelian braid statistics~\cite{Moore91,Read00,Read96}. 

Intense effort has been expended into testing various aspects of this proposal. Convincing evidence exists that the 5/2 FQHE derives from the CF Fermi sea: the CF Fermi sea has been seen at $\nu=5/2$ at somewhat elevated temperatures where the FQHE state is no longer present~\cite{Willett02}, and also at low temperatures in the close vicinity of filling factor $\nu=5/2$~\cite{Hossain18b}. A measurement of the non-Abelian statistics has been sought in interference experiments~\cite{Willett09,Willett13} that test certain theoretical predictions~\cite{DasSarma05,Stern06}, and also through the thermal Hall effect ~\cite{Banerjee18}. Theoretically, the MR ground state wave function has been found to provide a reasonable approximation for the exact Coulomb wave function~\cite{Morf98}, and also shown to be a better variational state than the CF Fermi sea~\cite{Park98b}. A recent work has constructed a BCS wave function for CFs and shown a p-wave pairing instability at $\nu=5/2$ but none at $\nu=1/2$~\cite{Sharma21}. However, a convincing theoretical demonstration of non-Abelian statistics for the excitations of the Coulomb interaction has not been possible so far.

A crucial theoretical development in this context has been a rigorous demonstration of non-Abelian braid statistics of the quasi{\em holes} (QHs) for a model three-body interaction in the LLL, denoted $\hat{V}^\text{Pf}_3$~\cite{Greiter91,Moore91,Nayak96,Read96,Tserkovnyak03}, which  obtains the MR ground state as well as QH wave functions as the exact zero energy states~\cite{Greiter91}. (This interaction is a generalization of previous two-body model interactions for the Laughlin state~\cite{Haldane83,Trugman85}.) For this model, the quasiholes are non-interacting. Besides a proof of principle for non-Abelian braid statistics, this provides a starting point from which one can hope to establish non-Abelian braid statistics for the QHs of the Coulomb interaction through adiabatic continuity as the $\hat{V}^\text{Pf}_3$ interaction is continuously deformed into the Coulomb interaction. 

An analogous demonstration of non-Abelian braid statistics for the quasi{\em particles} (QPs) for any model interaction has been missing, however. The $\hat{V}^\text{Pf}_3$ interaction does not lend itself to an exact solution for QPs, and its numerical solutions do not bring out a well separated a low-energy band of QP states that is consistent with the expectation from QPs with non-Abelian braid statistics\cite{Sreejith11b,Rodriguez12b}. 
In this article, we construct a model Hamiltonian, defined in terms of two- and three-body Haldane pseudopotentials\cite{Haldane83}, that produces exact solutions also for QPs, which are degenerate (that is, the quasiparticles are non-interacting, as is the case for quasiholes) and separated from other states by a gap. Our results for the QP spectrum are consistent with the prediction of the bipartite CF (BCF) model with exact LLL projection, as considered by Rodriguez {\em et al.} \cite{Rodriguez12b}. This model further allows exact solutions for the topological as well as the ordinary exciton. (The system with an odd number of composite fermions, i.e. with an unpaired CF, contains a topological exciton\cite{Sreejith11b}; it has been referred to in the literature as a neutral fermion also~\cite{Moller11}.) We also demonstrate, for finite systems accessible to our exact diagonalization (ED) study, that a path can be identified from our model to the $\hat{V}^\text{Pf}_3$ interaction along which both the ordinary and the topological neutral excitons evolve continuously, without any level crossings.

We briefly outline the motivation for our model here. Our starting point is a re-expression of the MR Pfaffian wave function as
 \be
 \Psi^\text{Pf}_{1/2}=
\mathcal{A}[\prod_{j<k}(z_j-z_k)^3 \times \prod_{j<k}(w_j-w_k)^3  \times  \prod_{j,k}(z_j-w_k)],
\label{eq:CauchyMR}
 \ee
where we have divided the $N$ particles into two partitions of $N/2$ particles each and denoted their positions by $\{z_1,\cdots z_{N/2} \}$ and $\{w_1,\cdots w_{N/2} \}$, $w_j\equiv z_{j+N/2}$, with $j,k=1,\cdots, N/2$. 
We have defined $z_j=x_j-iy_j$,
 assumed even $N$, set the magnetic length to unity, and suppressed the gaussian factor. The symbol $\mathcal{A}$ refers to antisymmetrization with respect to all of the coordinates. The equality of this wave function and the MR wave function follows from a Cauchy identity. The wave function inside the square brackets $[\cdots ]$ is the so-called Halperin 331 bilayer state~\cite{Halperin83}. In the spirit of the CF theory for bilayer states~\cite{Scarola01b}, we rewrite it as 
 \begin{widetext}
  \be
 \Psi^\text{Pf}_{1/2}=
\mathcal{A}[\prod_{j<k}(z_j-z_k)^2\Phi_1(\{ z_j\}) \times  \prod_{j<k}(w_j-w_k)^2\Phi_1(\{ w_j\}) \times \prod_{i,j}(z_i-w_j)], 
 \ee
 \end{widetext}
where $\Phi_1$ is the wave function of one filled LL and $\prod_{j<k}(z_j-z_k)^2\Phi_1(\{ z_j\})$ is the CF representation of the Laughlin wave function. A wave function with this structure, i.e. $\mathcal{A}[ \Psi^{\rm CF}(\{ z_j\}) \Psi^{\rm CF}(\{ w_j\})\prod_{i,j}(z_i-w_j)^m]$, is referred to as the bipartite CF, or BCF wave function. (The corresponding bilayer wave functions represent two-component composite fermions that bind $2p$ intra-layer flux quanta and $m$ inter-layer flux quanta. These are denoted by $^{2p}_m$CFs~\cite{Li19}. Many bilayer CF states have been observed in double quantum wells and double layer graphene~\cite{Eisenstein92,Suen92,Li17,Li19,Liu19}. Their antisymmetrization produces single layer states, which are relevant to the present work.)

An advantage of the BCF form is that it can be generalized to other fractions by replacing $\Phi_1$ by $\Phi_n$~\cite{Hermanns10,Sreejith11b}. 
More relevant to our present work is the observation that this form allows a construction of candidate wave functions for QPs, QHs and neutral excitations, which correspond to analogous (but known) excitations at $\nu=1$ in the factors $\Phi_1$. Consider, for example, the state obtained by adding $n$ flux quanta to the system, which corresponds to creating $n$ QHs in each factor of $\Phi_1$, and thus contains a total of $2n$ QHs. 
The basis for these QHs can be straightforwardly constructed. The BCF description of QHs can be shown to be equivalent to the MR QHs in the Pfaffian construction, again using the Cauchy identity. The origin of non-Abelian statistics lies in the fact that simply identifying the positions $\{\eta_1,\eta_2,\dots \eta_{2n}\}$ of the QHs does not uniquely fix the wave function, because there are several ways of distributing the QHs into the two partitions. One would expect a 
degeneracy of $(2n)!/[2 (n!)^2]$ for $2n$ QHs with fixed locations.
It turns out that because of the non-trivial linear dependencies of these basis functions~\cite{Nayak96,Read96,Bonderson12}, the actual degeneracy is $2^{n-1}$, which is crucial for identifying the QHs with Majorana zero modes. 

We note here the conceptual, and perhaps also practical, significance of the availability of an exactly solvable model for the QH states. The 
exact linear dependencies of the QH basis functions are special to the MR wave functions. Though one expects that slight perturbations should not change the physics, they can eliminate linear dependencies. In such a case, a simple counting of linearly independent basis functions would miss the underlying physics.
It is therefore conceptually useful to have the model interaction $\hat{V}^\text{Pf}_3$, which produces the MR ground state and QH wave functions as exact zero energy states, separated from excited states by a gap. The counting of zero energy QH states of this model is consistent with the expectation with non-Abelian statistics. Adding a weak perturbation to the Hamiltonian would produce eigenstates different from the MR QHs, but the physics of the MR model would remain valid so long as the QH states evolve adiabatically.

The BCF formulation also allows a construction of candidate wave functions for QPs~\cite{Sreejith11b,Rodriguez12b} (a conformal field theory based construction of the non-Abelian QP states is given in Refs.~\cite{Hansson09b,Hansson17}). The BCF state with $2n$ QPs is obtained by replacing $\Phi_1$ by a state $\Phi^{\rm n-qp}_1$ in which the LLL is full and there are $n$ electrons in the second LL:
\begin{widetext}
\be
 \Psi^\text{BCF, 2n-qp}_{1/2}=
{\cal P}_{\rm LLL}\mathcal{A}[\prod_{j<k}(z_j-z_k)^2\Phi_1^{\rm n-qp}(\{ z_j\})\times \prod_{j<k}(w_j-w_k)^2\Phi_1^{\rm n-qp}(\{ w_j\})\times \prod_{j,k}(z_j-w_k)].
\label{2nqp}
\ee
\end{widetext}
The number of linearly independent QP states depends on how the LLL projection is performed. Ref.~\cite{Sreejith11b} used the Jain-Kamilla method for the LLL projection\cite{Jain97,Jain97b} and found that the QP states do not show any linear dependencies. This would suggest that the quasiparticles may behave differently from quasiholes. In an insightful work, Rodriguez {\em et al.}\cite{Rodriguez12b} showed that if the LLL projection of the BCF wave functions is performed slightly differently (namely the ``exact'' projection, as opposed to the Jain-Kamilla projection), then there exist exact linear dependencies  for the QP wave functions too. (As the exact and Jain-Kamilla projected wave functions are very close, approximate linear dependencies are expected for the latter as well. We have explicitly checked this through singular value decomposition of the overlap matrix.)
Ref.~\cite{Rodriguez12b} found that counting of the quasiparticles is the same as that of the quasiholes except in the smallest few angular momentum sectors, and concluded that edge physics of both the QP and QH states are essentially consistent with the expectation from the conformal field theory of MR Pfaffian state.

Independently of how these wave functions are projected into the LLL, these wave functions are not exact eigenstates of the $\hat{V}^\text{Pf}_3$ interaction model. The spectrum of the $\hat{V}^\text{Pf}_3$ interaction obtained by ED does not exhibit a well separated band of states that can be identified with QP states.

\begin{figure}[t]
 \begin{center}
  \includegraphics[width=\columnwidth]{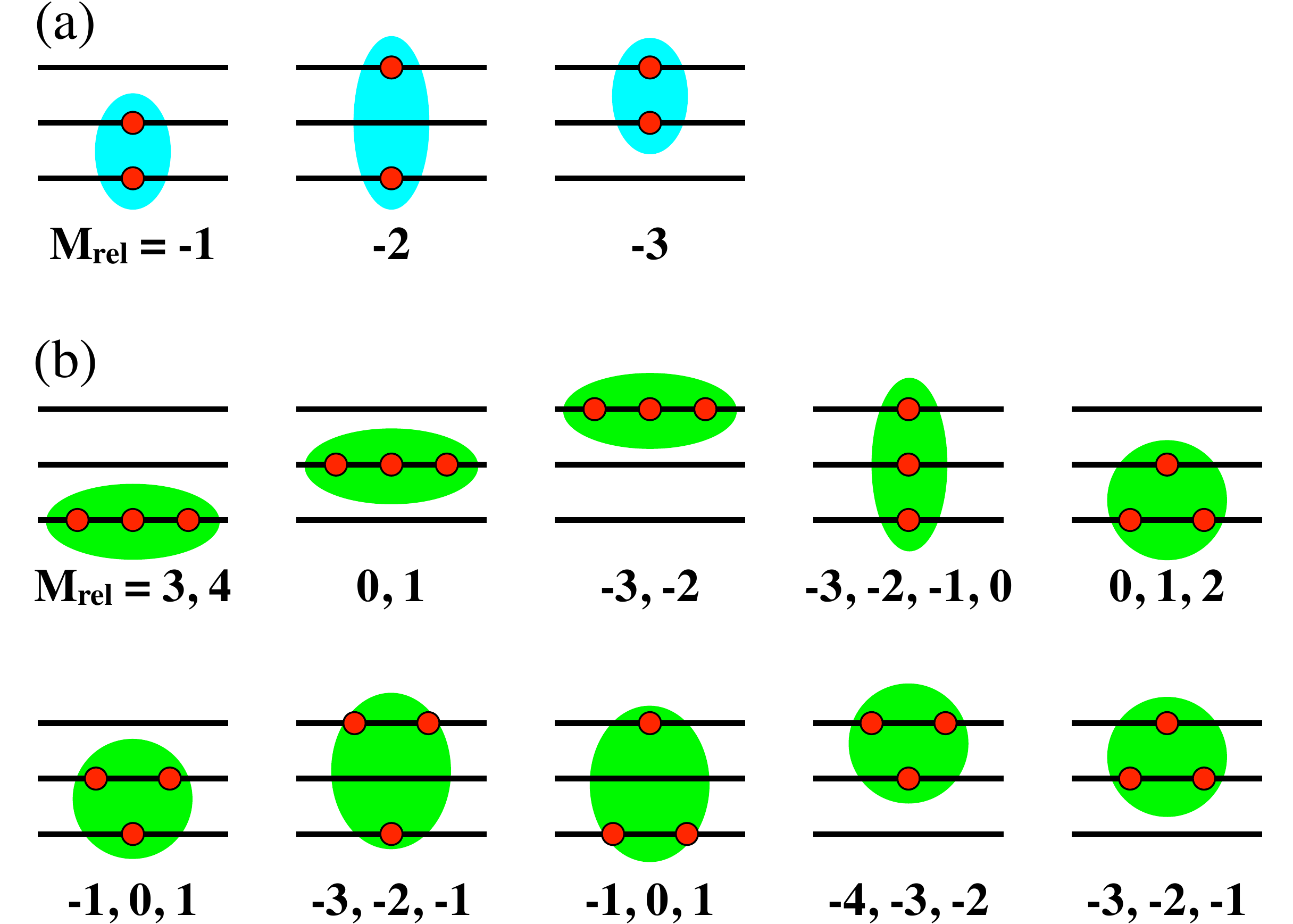}
 \end{center}
 \caption{This figure shows all of the (a) pairs and (b) triplets residing within the lowest three LLs that are absent in wave functions of the BCF form. The integers in each panel are the relative angular momenta of these pairs / triplets. The model interaction penalizes these configurations.\label{fig:pictorial}}
\end{figure}

This background motivates us to construct a reference Hamiltonian for QP
states, analogous to $\hat{V}^\text{Pf}_3$ for the QH problem, 
for which the QPs are noninteracting, and thus the QP states are unambiguously identifiable.
That is the objective of the present work. Motivated by the strategy in Anand {\em et al.}\cite{Anand21} we proceed as follows:

\begin{itemize} 

\item We do not confine to the LLL but allow higher LLs. We introduce an interaction in terms of generalized intra- as well as inter-LL Haldane pseudopotentials. This interaction has two- and three-body terms and penalizes pairs and triplets that are absent in the BCF form, shown pictorially in Fig.~\ref{fig:pictorial} for electrons occupying the lowest three LLs. The general form of the interaction, including arbitrary LLs, is given below.

\item We construct the interaction such that it preserves the LL index (i.e., does not alter the LL occupation). It thus commutes with the kinetic energy part of the Hamiltonian.

\item Finally, we take the limit where the interaction energy is much larger than the cyclotron energy $\hbar \omega_c$. The low-energy spectrum thus consists of states that have zero interaction energy and are eigenstates of the kinetic energy. These form bands  separated by an integer multiple of the cyclotron energy. This allows an unambiguous identification of the QP states as the degenerate states with the lowest kinetic energy.

\end{itemize}

The important question is if we can understand the ``counting'' of the low energy states which allows us to infer the braid statistics of the QPs.
To this end, we can compare the counting of the QP states in three reference systems: (i) 
the exact spectrum of the model interaction; (ii) 
the exactly projected BCF wave functions~\cite{Rodriguez12b}; and (iii) 
the ``equivalent'' QH systems (defined below).
In every case that we could test, we find that the counting of QP states of the model is identical to the counting predicted by the BCF model defined through an exact LLL projection\cite{Rodriguez12b}.
A non-trivial aspect of the model is that the actual number of degenerate QP states is in general (for four or more QPs) smaller than the apparent expectation from the BCF model, indicating exact linear dependencies for the QP basis functions (as seen previously for QH basis functions), even though the QP wave functions are much more complicated than the QH wave functions. 
The equivalent QP and QH systems also match in their edge behaviors as indicated by the agreement between the counting of QH and the counting of QP states at large angular momenta.

This model also allows for exact solutions for the ordinary neutral excitation, which corresponds to a particle hole excitation in one of the $\Phi_1$ factors. The system with an odd $N$ has one unpaired CF (or a ``neutral fermion''). In the BCF formulation, this maps into a particle hole pair in which the particle and the hole reside in different $\Phi_1$ factors; this has been called a topological exciton. We show below that both the ordinary and the topological excitons evolve continuously, without gap closing, as we deform our model Hamiltonian into the LLL $\hat{V}^\text{Pf}_3$ interaction.

It ought to be stressed that even though our model is inspired by the BCF structure, its QP solutions are {\em not} given by wave functions of the type in Eq.~\eqref{2nqp}, because the projected BCF wave functions do not involve higher LLs whereas the unprojected BCF wave functions are in general not eigenstates of the kinetic energy.
As in Ref.~\onlinecite{Anand21}, we can formally write the wave functions for $2n$ QPs that have zero interaction energy:
\begin{widetext}
  \be
 \Psi^\text{2n-qp}_{1/2}=
\mathcal{A}[\prod_{j<k}(\hat{Z}_j-\hat{Z}_k)^2\times\prod_{j<k}(\hat{W}_j-\hat{W}_k)^2\times\prod_{i,j}(\hat{Z}_i-\hat{W}_j) \times \Phi^\text{n-qp}_1(\{ z_j\})\Phi^\text{n-qp}_1(\{ w_j\})], 
 \ee
 \end{widetext}
where $\hat{Z}_j$ and $\hat{W}_j$ are guiding center operators. As the guiding center coordinates commute with the kinetic energy, these wave functions are eigenstates of the kinetic energy with eigenvalue $2n$ in units of the cyclotron energy. These wave functions are not amenable to explicit evaluation, however. Analogous wave functions can be constructed for all states in the zero interaction energy sector. We conjecture that all zero interaction energy eigenstates have this form.

The paper is organized as follows. In Sec.~\ref{sec:background}, we review the MR wave function, the BCF wave function, and an exact model for Abelian FQH states. In Sec.~\ref{sec:modelHam}, our solvable model and its exact solutions are constructed. In Sec.~\ref{sec:num}, we clarify the non-Abelian braiding  statistics of QPs by showing that the degeneracy of QP states is in agreement with that of QH states at large angular momenta. In Sec.~\ref{sec:fcharge}, we provide numerical evidence of the fractional charge of QPs by using our exact wave functions. In Sec.~\ref{sec:adiabatic}, we demonstrate adiabatic continuity for the ordinary exciton and the topological exciton as we deform our model Hamiltonian into the LLL $\hat{V}_3^\text{Pf}$ interaction.

\section{Background}
\label{sec:background}
\subsection{Moore Read Pfaffian wave function}

We begin by reviewing the MR Pfaffian state and the
multi-fold degeneracy of its quasiholes leading to non-Abelian excitations. The
$\nu=1/2$ 
MR Pfaffian wave function, for even particle number $N$, 
is defined as~\cite{Moore91}
\begin{align}
 \Psi^\text{Pf}_{1/2}
 =\text{Pf}\left(\frac{1}{z_i-z_j}\right)\prod_{i<j}^N\left(z_i-z_j\right)^2.
 \label{eq:MRwf}
\end{align}
This is identical to Eq.~\eqref{eq:CauchyMR} through a Cauchy identity. An 
excited state 
$\prod_i(z_i-\eta)\Psi^\text{Pf}_{1/2}$, generated by piercing a flux quantum 
at $\eta$, has a vortex with
a local charge of $e/2$. Since the MR wave function describes a 
paired state of CFs~\cite{Jain89,Read00}, it can support half quantum vortices of charge $e/4$. These are the quasihole excitations of the system. A state with two such quasiholes located at $\eta_1$ and $\eta_2$  
is given by the wave function 
\begin{equation}
 \Psi^\text{Pf,2-qh}_{1/2}=\text{Pf}
  \left(
   \frac{(z_i-\eta_1)(z_j-\eta_2)
   +(i\leftrightarrow j)}{z_i-z_j}
  \right)
  \prod_{i<j}^N\left(z_i-z_j\right)^2.
\end{equation}
This formulation can be generalized to the case of $2n$ QHs:
\begin{align}
\Psi^\text{Pf,2n-qh}_{1/2}
&=\text{Pf}\left(M_{ij}\right)
\prod_{i<j}^N\left(z_i-z_j\right)^2,
\label{eq:MRwf_2n-qh}
\end{align}
where
\begin{align}
 M_{ij}=
 \frac{\prod_{\alpha=1}^n(z_i-\eta_\alpha)(z_j-\eta_{n+\alpha})
 +(i\leftrightarrow j)}{z_i-z_j}.
\label{eq:Mij}
\end{align}

The wave functions $\Psi^\text{Pf}_{1/2}$ and $\Psi^\text{Pf,2n-qh}_{1/2}$ are
zero energy eigenfunctions of a Hamiltonian with a short-range three-body 
interaction~\cite{Greiter91,Read96}:
$H=\sum_{i<j<k}\hat{V}_3^\text{Pf}(i,j,k)$ with
\begin{align}
 \hat{V}_3^\text{Pf}
 =VP_{ijk}(M_\text{min}),
 \label{eq:V_3}
\end{align}
where $P_{ijk}(M_\text{rel})$ is the projection onto a three particle state of 
relative angular momentum $M_\text{rel}$ (we will use the symbol $M_{\rm rel}$ 
for the relative angular momentum) and $M_\text{min}$ is $3$ here, 
corresponding to the closest approach of three particles. 

While $\Psi^\text{Pf}_{1/2}$ is the unique zero mode at exactly $\nu=1/2$, 
the QH states are generally degenerate. 
The wave function $\Psi^\text{Pf,2n-qh}_{1/2}$ has $2n$ QHs at $\eta_1,\ldots,\eta_{2n}$ and in the construction of the wave function, these are partitioned into two groups of $n$ each as 
$(\eta_1,\ldots,\eta_n)$ and 
$(\eta_{n+1},\ldots,\eta_{2n})$ as shown in Eq.~\eqref{eq:Mij}.
Due to the multiple ways of making this partition, we can write $(2n)!/[2(n!)^2]$ wave 
functions, all of them having QHs located at  $(\eta_1,\ldots,\eta_{2n})$. 
Remarkably, however, only $2^{n-1}$ of these are linearly independent~\cite{Moore91,Nayak96,Read96}.

This linear dependence can be associated with $2n$ Majorana modes on each QH. A collection of $2n$ Majorana fermions can be combined into $n$ complex fermions, each of which can be occupied or unoccupied, resulting in a dimension of $2^n$ for the space of QHs. However, the parity of the fermions is conserved, which reduces the dimension to 
$2^{n-1}$~\cite{Nayak96,Read96,Ivanov01}. 
This degenerate subspace gives an irreducible representation of the braid group, resulting in the non-Abelian braid statistics.
\subsection{Bipartite CF description}

\begin{figure}
\includegraphics[width=\columnwidth]{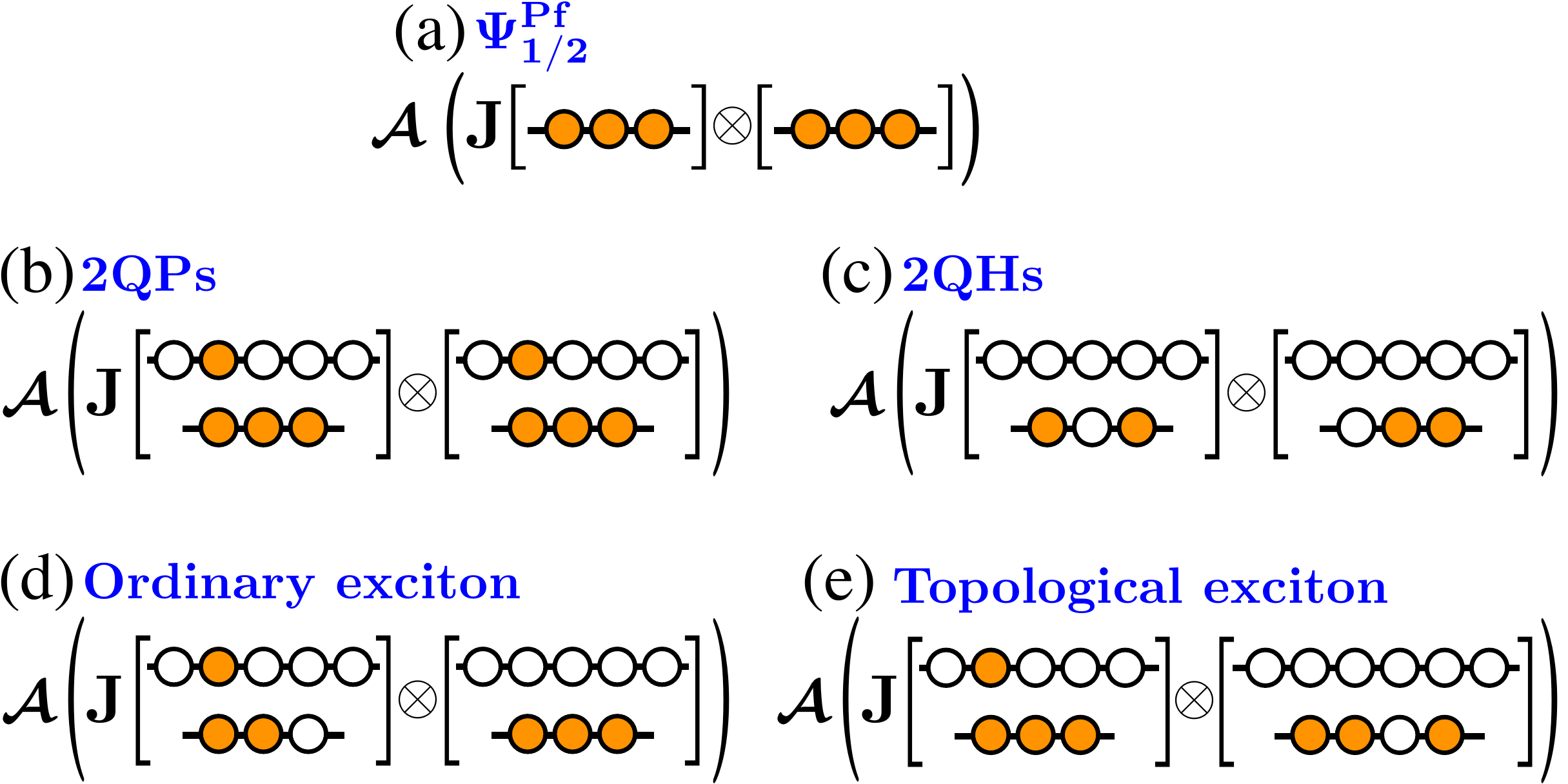}
\caption{BCF wave function of (a) the ground state $\Psi^{\rm Pf}_{1/2}$ and (b)-(e) various excitations 
 at filling fraction $1/2$ is shown schematically. 
 The pictures inside the square brackets depict the 
 particle occupancies of Slater determinant states $\Phi_1^\alpha$ in each 
 partition, $J$ represents the Jastrow factor defined in 
 Eq.~\eqref{eq:J(z_j,w_k)}, and ${\cal A}$ refers to antisymmetrization. (The LLL projection operator ${\cal P}_{\rm LLL}$ is not shown explicitly.)
}
\label{fig:schematic_bcf}
\end{figure}

The BCF description proposes a set of wave functions for incompressible states and their excitations at a sequence of filling fractions $\nu={n}/{[(2p+1)n\pm 1]}$, where $n$ and $p$ are integers~\cite{Sreejith11b}. The BCF wave functions have the general form
\begin{equation}
\Psi^{\rm BCF,(\alpha_1,\alpha_2)}_{\frac{n}{(2p+1)n\pm 1}} = \mathcal{A}[\Psi^{\rm CF,\alpha_1}_{\frac{n}{2pn\pm 1}}(\{z_i\})\Psi^{\rm CF, \alpha_2}_{\frac{n}{2pn\pm 1}}(\{w_j\})\prod_{i,j}(z_i-w_j)],\label{eq:BCFwf}
\end{equation}
where the factor 
\begin{align}
 \Psi_{n/(2pn\pm 1)}^{\rm CF,\alpha}
 =\mathcal{P}_\text{LLL}\Phi_{\pm n}^\alpha(\{z_i\})\prod_{j<k}(z_j-z_k)^{2p}
 \label{eq:CFwf}
\end{align}
is the Jain CF wave function. $\Phi_{+n}^\alpha$ is a Slater determinant 
state of single particle Landau eigenstates at filling $n$, where
$\alpha$ labels the distinct states and
$\Phi_{-n}^\alpha\equiv (\Phi_n^\alpha)^*$. For the incompressible state, 
$\Phi_{\pm n}^\alpha$ is the ground state completely occupying the $n$th LLs.

Different BCF excited states labeled by $(\alpha_1,\alpha_2)$ in Eq.~\eqref{eq:BCFwf} have excited states labeled by $\alpha_1$ and $\alpha_2$ of the Jain CF states in the two partitions. A pair of QPs/QHs of the BCF state corresponds to one QP/QH each in the two CF states formed in the two partitions.
Each BCF QP/QH have charges $\nicefrac{1}{[(2p+1)n+1]}$ and the incompressible states occur at a shift of $\pm (n+2p)$ on the spherical geometry. 
Note that though the BCF wave function is constructed in terms of product of two wave functions each containing only one part of the full number of particles, the wave function obtained at the end of antisymmetrization is a fully valid many identical fermion wave function.

For later convenience, let us rewrite
Eq.~\eqref{eq:BCFwf}
with $n=1,p=1$, occurring at a filling fraction $\nu=1/2$, as
\begin{equation}
 \Psi^{\rm{BCF},(\alpha_1,\alpha_2)}_{1/2}
  =\mathcal{A}\left[\mathcal{P}_\text{LLL} J(\{z_j\},\{w_j\})
   \Phi^{\alpha_1}_1(\{z_j\})\Phi^{\alpha_2}_1(\{w_j\})
  \right],
  \label{eq:BCFwf_J}
\end{equation}
where 
\begin{align}
 J
 \equiv 
 \prod_{j<k}(z_j-z_k)^2\prod_{j<k}(w_j-w_k)^2\prod_{j,k}(z_j-w_k).
 \label{eq:J(z_j,w_k)}
\end{align}
The incompressible state is identical to the MR wavefunction 
$\Psi^\text{Pf}_{1/2}$. 
Its bulk QHs and QPs can be created only in pairs:
$\Psi^\text{BCF,2n-qh}_{1/2}$ and $\Psi^\text{BCF,2n-qp}_{1/2}$ are 
constructed from $\Phi_1^\text{n-qh}\Phi_1^\text{n-qh}$ or
$\Phi_1^\text{n-qp}\Phi_1^\text{n-qp}$ via Eq.~\eqref{eq:BCFwf_J}. 
These are schematically shown in Figs.~\ref{fig:schematic_bcf}(a)-(c).
$\Psi^\text{BCF,2n-qh}_{1/2}$ are identical to 
the MR QH states $\Psi^\text{Pf,2n-qh}_{1/2}$ and thus have the linear 
dependencies among them. $\Psi^\text{BCF,2n-qp}_{1/2}$ 
also show either approximate or exact linear dependencies depends on how they are projected into the LLL~\cite{Rodriguez12b}.

There are two possible ways to construct neutral excitations of the state [Figs.~\ref{fig:schematic_bcf}(d) and (e)]. We can create a QP-QH pair in either one of the partitions, which produces an ordinary neutral exciton. Alternatively, if the QP and QH reside in different partitions, we get what is variously known as a topological exciton, an unpaired CF, or a neutral fermion. 
{(This is referred to as topological exciton below.)}

In the spherical geometry, the incompressible state for the short-range three-body interaction occurs at a flux $2Q=2N-3$, when the number of particles is even. The incompressible state and its ordinary neutral excitations are well approximated by the BCF description. When $N$ is odd, however, the spectrum instead has a low energy dispersing mode rather than an incompressible state. This dispersing mode is well described by the topological exciton mode. 
Note that the topological exciton mode has odd number of particles; consistency of the topological exciton wave function demands that the number of particles in the partition with the QP has one additional particle compared to the partition that holds the QH excitation \cite{Sreejith11}.

\subsection{Model Hamiltonian for $\nu=n/(2pn+1)$ FQH states}
In this section, we present a review of the exactly solvable model introduced in Ref.~\onlinecite{Anand21} for Abelian FQH  states at Jain fractions $\nu=n/(2pn+1)$. The model Hamiltonian has the form 
\begin{equation}
\hat{H}=\sum_j\hat{\bm{\pi}}_j^2/2m+\sum_{i<j}\hat{V}(i,j),
\end{equation}
where $\hat{\bm{\pi}}$ is the kinetic momentum and $\hat{V}$ is the model interaction between two particles defined below. This model produces the energy spectrum of non-interacting CFs when the interaction is infinitely larger than the cyclotron energy.

The interaction $\hat{V}$ is constructed so that an analog of the Jain CF wave functions is a zero-energy eigenfunction. 
The standard Jain wave function is constructed by multiplying the integer quantum Hall (IQH) state $\Phi^\alpha_{n}$ by the Jastrow factor $\prod_{j<k}(z_j-z_k)^{2p}$ and projecting it into the LLL, as shown in Eq.~\eqref{eq:CFwf}. This Jastrow 
factor increases the relative angular momentum $M_\text{rel}$ of each electron 
pair in $\Phi^\alpha_{n}$ by $2p$.

\begin{figure}[t!]
\begin{center}
\includegraphics[width=\columnwidth]{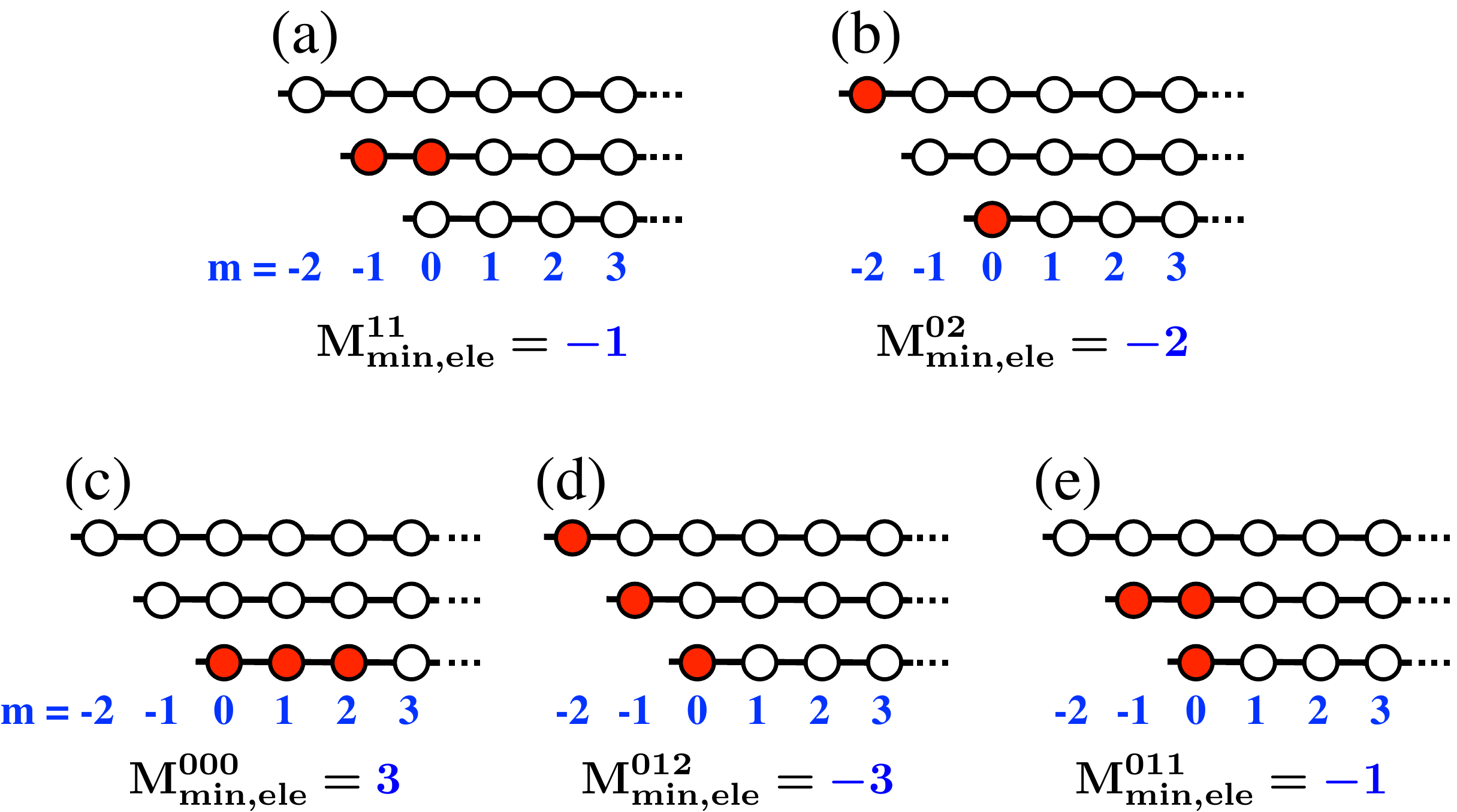}
\end{center}
 \caption{This figure shows the minimum relative angular momentum for (a)(b) electron pairs as well as (c)-(e) electron triplets for certain LL occupations. $M_{\rm min,ele}^{n_1n_2}$ is the minimum relative angular momentum for two electrons in $n_1^{\rm th}$ and  $n_2^{\rm th}$ LLs, and $M_{\rm min,ele}^{n_1n_2n_2}$ is the minimum relative angular momentum for three electrons in $n_1^{\rm th}$, $n_2^{\rm th}$ and $n_3^{\rm th}$ LLs. The y-axis shows LLs, labeled $n=0, 1, \cdots$; the x-axis labels the single particle angular momentum $m=-n, -n+1, \cdots$.}
 \label{fig:Mmin_Full}
\end{figure}
Keeping this in mind, let us consider two electrons in the $n_1$th and $n_2$th 
LLs. Since the $n$th LL can have single particle angular momenta $m=-n,-n+1,\ldots$, the minimum $M_\text{rel}$ of the pair is
$M_\text{min,ele}^{n_1n_2}=-n_1-n_2+\delta_{n_1n_2}$, as shown in Figs.~\ref{fig:Mmin_Full}(a) and (b). Electron pairs with relative angular momentum smaller than $M_\text{min,ele}^{n_1n_2}$ are disallowed by the Pauli principle.
Converting electrons into composite fermions increases $M_{\rm rel}$ of each pair by $2p$, 
while at the same time preserving their LL indices.
The smallest $M_{\rm rel}$ in the resulting state is $M_\text{min,CF}^{n_1n_2}=M_\text{min,ele}^{n_1n_2}+2p$, i.e., pairs with $M_{\rm rel}<M_\text{min,CF}^{n_1n_2}$ are absent. A model interaction is 
introduced to penalize only these absent pairs:
\begin{align}
 \hat{V}=\sum_{n_1,n_2=0}^{\infty}
 \sum_{M_{\rm rel}=M_\text{min,ele}^{n_1n_2}}^{M_\text{min,CF}^{n_1n_2}-1}
 V_{M_{\rm rel}}^{n_1n_2}P_{n_1n_2}(M_{\rm rel}),
 \label{eq:AJS-V}
\end{align}
where $P_{n_1n_2}(M_{\rm rel})$ is the projection operator on states of a pair from the 
$n_1$th and $n_2$th LLs with relative angular momentum $M_{\rm rel}$ and $V_{M_{\rm rel}}^{n_1n_2}$
is a generalization of the Haldane's pseudopotentials. 

The interaction conserves the LL index, i.e., the particle number in each LL is preserved.
It does not yield the unprojected Jain CF state as the eigenstate because it is not an eigenstate of the kinetic energy and mixes basis functions with different LL occupancies. 
 The following set of CF-like wave functions, where the coordinates in the Jastrow factor are replaced by LL preserving guiding center operators, are zero interaction energy eigenstates:
\begin{align}
 \Psi^\text{$\alpha$}_{\nu^*/(2p\nu^*+1)}
 =\prod_{j<k}(\hat{Z}_j-\hat{Z}_k)^{2p}\times\Phi_{\nu^*}^\alpha,
 \label{eq:AJS-psi}
\end{align}
where $\hat{Z}=\hat{z}-i(\hat{\pi}_x-\hat{\pi}_y)/\hbar$ is the guiding center coordinate and $\Phi_{\nu^*}^\alpha$ is a Slater determinant state at filling 
fraction $\nu^*$.
This follows because the Jastrow operator 
$\prod_{j<k}(\hat{Z}_j-\hat{Z}_k)^{2p}$ increases $M_{\rm rel}$ of every pair by $2p$ without changing the LLs. Thus each pair of particles has $M_{\rm rel}$ greater than $M^{n_1n_2}_{\rm min, CF}$. 
Numerically, it could be demonstrated that linearly independent basis functions $\Phi_{\nu^*}^\alpha$ produce linearly independent states $\Psi^\text{$\alpha$}_{\nu^*/(2p\nu^*+1)}$, which, furthermore, 
span the full zero energy eigenspace of the interaction \cite{Anand21}. Thus there is a one-to-one correspondence between states $\Phi_{\nu^*}^\alpha$ at filling $\nu^*$ and the zero interaction energy states $\Psi^\text{$\alpha$}_{\nu^*/(2p\nu^*+1)}$. Incompressible states are obtained for $\nu^*=n$.

If we take the strong interaction limit, $V_{M_{\rm rel}}^{n_1n_2}/\hbar\omega_c\rightarrow\infty$, all other states, which have finite interaction eigenvalues, are projected out. The degeneracy of the zero interaction energy space is lifted by the kinetic energy $\sum_j\hat{\bm{\pi}}_j^2/2m$.
The guiding center $\hat{Z}_k$, and therefore the guiding center Jastrow factor $\prod_{i<j}(\hat{Z}_i-\hat{Z}_j)^{2p}$, commutes through the kinetic energy; we see that the kinetic energy of each zero interaction energy state [Eq.~\eqref{eq:AJS-psi}] is same as that of the Slater determinant $\Phi_{\nu^*}^\alpha$.
Thus the spectrum of $\hat{H}$ at $\nu=\nu^*/(2p\nu^*+1)$ is the same as that of non-interacting particles at filling fraction $\nu^*$.

\section{Model Hamiltonian for non-Abelian state at $\nu=1/2$}

\label{sec:modelHam}

In this section, we construct a model Hamiltonian for the $\nu=1/2$ state 
by using a strategy similar to that in the previous section, which gives not only the ground state 
and the QHs, but also QPs, neutral excitations, and all other excitations. The BCF representation would 
be the key to our construction. As one may anticipate, the exact solutions are given by wave functions of the form
\begin{equation}
\Psi_{1/2}^{(\alpha_1,\alpha_2)} = \mathcal{A}[\hat{J} \Phi_1^{\alpha_1}({z_j})\Phi_1^{\alpha_2}({w_j})],
 \label{eq:KSSJ-psi}
\end{equation}
which are obtained by replacing the Jastrow factor $J$ with an operator $\hat{J}$ in the BCF wave function in 
Eq.~\eqref{eq:BCFwf_J}, where 
\begin{align}
 \hat{J}&\equiv J(\{\hat{Z}_j\},\{\hat{W}_j\})\non
 &=
 \prod_{j<k}(\hat{Z}_j-\hat{Z}_k)^2\prod_{j<k}(\hat{W}_j-\hat{W}_k)^2\prod_{j,k}(\hat{Z}_j-\hat{W}_k).
 \label{eq:hatJ(z_j,w_k)}
\end{align}
Note that the action of $\hat{J}$ on $\Phi^{\alpha_1}_1\Phi^{\alpha_2}_1$ does not affect the LL indices of the particles in the Slater determinants. Thus we can associate well-defined LL occupancies to the particles in the full wavefunction $\Psi_{1/2}^{(\alpha_1,\alpha_2)}$. 

\begin{figure}[t!]
\begin{center}
\includegraphics[width=\columnwidth]{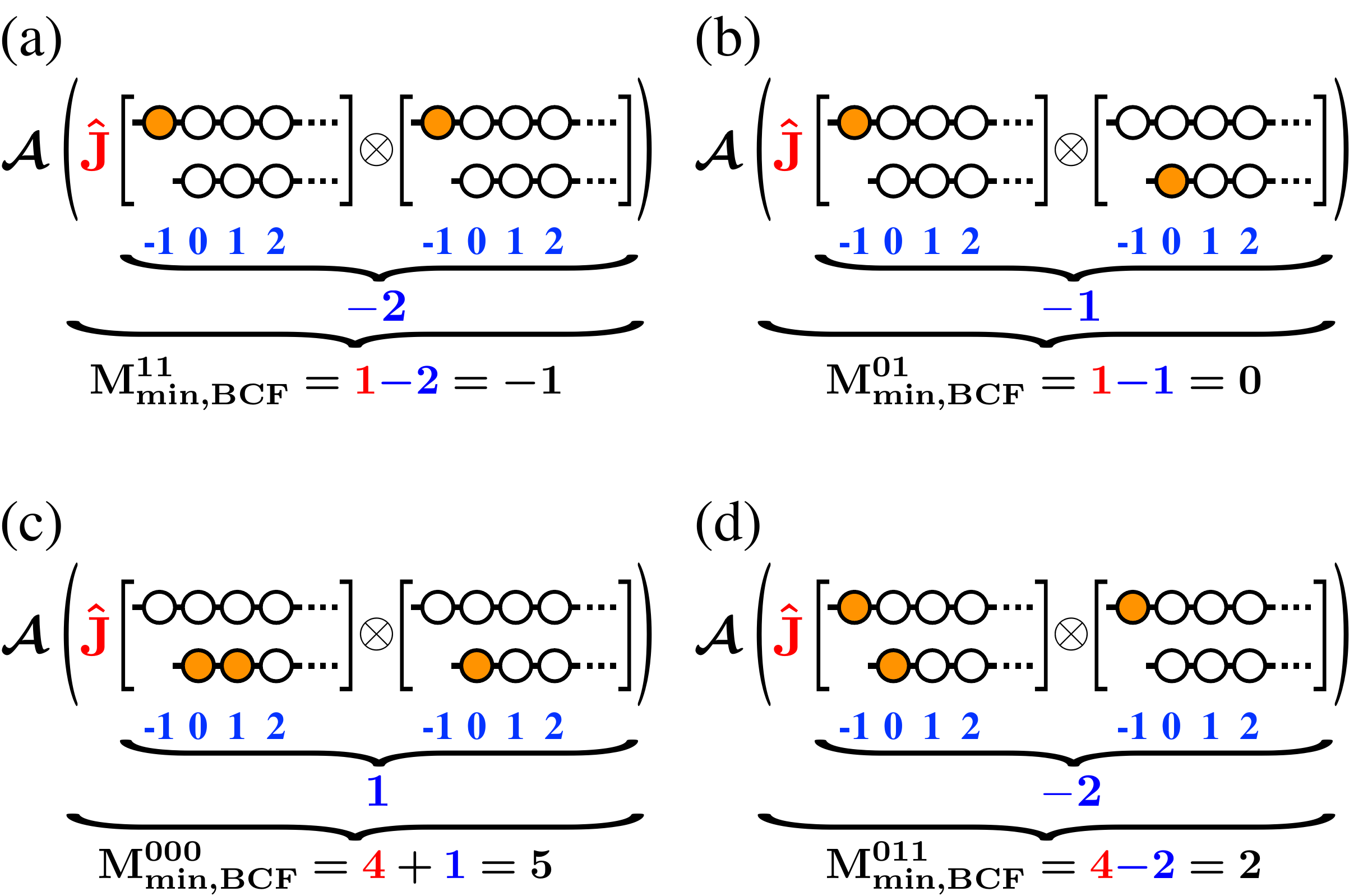}
\end{center}
 \caption{This figure pictorially explains the calculation of the minimum relative angular momentum in the BCF description for pairs and triplets with specified LL occupations.  $M_{\rm min,BCF}^{n_1n_2}$ is the minimum relative angular momentum for a pair in $n_1^{\rm th}$ and $n_2^{\rm th}$ LLs, and $M_{\rm min,BCF}^{n_1n_2n_2}$ is the minimum relative angular momentum for a triplets in $n_1^{\rm th}$, $n_2^{\rm th}$ and $n_3^{\rm th}$ LLs. The x-axis labels the single particle angular momentum $m$, the y-axis shows LLs, labeled $n=0,1$, and occupied orbitals are indicated by filled circles. The number (colored in blue) below the upper underbrace is the minimum angular momentum in the state $ \Phi_1^{\alpha_1}(\{z_j\})\Phi_1^{\alpha_2}(\{w_j\})$; the number below the lower underbrace is the minimum relative angular momentum in the full state, noting that the Jastrow operator $\hat{J}$ adds two (one) to the relative angular momentum of an intra (inter) partition pair. (The contribution of $\hat{J}$ to the relative angular momentum is shown in red.) The symbol ${\cal A}$ refers to antisymmetrization. The interaction in Eq.~\eqref{eq:KSSJ-H} imposes a penalty for all smaller relative angular momenta.}
 \label{fig:Mmin_bcf}
\end{figure}

We argue below that the suitable interaction Hamiltonian for which $\Psi_{1/2}^{(\alpha_1,\alpha_2)}$ are the zero interaction  energy states has the following form:
\begin{equation}
\hat{H} = \sum_i \frac{\hat{{\boldsymbol{\pi}}}^2_i}{2m} + \sum_{i<j}\hat{V}_2(i,j) + \sum_{i<j<k}\hat{V}_3(i,j,k).\label{eq:KSSJ-H}
\end{equation}
The interactions $\hat{V}_2$ and $\hat{V}_3$ are constructed in such a manner that they penalize pairs and triplets missing in the modified BCF wavefunctions $\Psi_{1/2}^{(\alpha_1,\alpha_2)}$.
The essential idea behind which pseudopotentials are included in the Hamiltonian is pictorially explained in Fig.~\ref{fig:Mmin_bcf}. Briefly, the total relative angular momentum of the pair / triplet gets contributions from the Slater determinants of the two partitions as well as from $\hat{J}$. Once we specify the LL indices of the electrons in the pair / triplet, the minimum relative angular momentum is obtained by distributing particles in a given LL into different partitions to the extent allowed by the Pauli principle. This allows minimization of the angular momentum contribution from the Slater determinants as well as from the Jastrow operator $\hat{J}$, which contributes 2 to intra-partition pairs but only 1 for inter-partition pairs. Once we have determined the minimum relative angular momentum in the BCF state, we construct an interaction Hamiltonian that imposes a penalty on all pairs / triplets with smaller relative angular momenta.

We now discuss this in greater detail.

\subsection{Two body component $\hat{V}_2$}
\label{sec:V2}
We begin with the construction of the interaction $\hat{V}_2$. 
Consider two filled orbitals from LLs $n_1$ and $n_2$ located in partitions $p_1$ and $p_2$ in the Slater determinant product $\Phi^{\alpha_1}_1\Phi^{\alpha_2}_1$. (The partition index $p_i$ can take values $1$ and $2$. $p_1=p_2$ if the two orbitals are in the same partition and $p_1\neq p_2$ if the orbitals are in different partitions). 
It can be seen, by inspection of individual cases, that its minimum possible $M_{\rm rel}$ in a general BCF wavefunction is given by $-n_1-n_2+\delta_{n_1n_2}\delta_{p_1p_2}$.
Action of $\hat{J}$ on $\Phi^{\alpha_1}_1\Phi^{\alpha_2}_1$ increases this by $1+\delta_{p_1p_2}$ to
\begin{equation}
 M_{\rm min}^{n_1,n_2}(p_1,p_2)
 =-n_1-n_2+\delta_{n_1n_2}\delta_{p_1p_2}+1+\delta_{p_1p_2}.
 \label{eq:M_minp1p2}
\end{equation}
In $\Psi_{1/2}^{(\alpha_1,\alpha_2)}$, the orbitals from LLs $n_1$ and $n_2$ can be in the same or different partitions; 
these cases are associated with different values of $M_{\rm min}^{n_1,n_2}(p_1,p_2)$.
Therefore the smallest possible $M_\text{rel}$ for such a pair in a general BCF wave function 
is given by 
the minimum of $M_\text{min}^{n_1n_2}(p_1,p_2)$ across all choices of $p_1,p_2$:
\begin{align}
 M_\text{min,BCF}^{n_1n_2}
 &=\min_{p_1,p_2}
 \left[
 M^{n_1n_2}_\text{min}(p_1,p_2)
 \right]\non
 &=-n_1-n_2+1.
\end{align}
[Figures~\ref{fig:Mmin_bcf}(a) and (b) show examples of pairs producing $M_{\rm rel}=M_\text{min,BCF}^{n_1n_2}$ channels.] Namely, pairs with $M_\text{rel}<M_\text{min,BCF}^{n_1n_2}$ are absent in $\Psi_{1/2}^{(\alpha_1,\alpha_2)}$.  
We construct $\hat{V}_2$ in such a way that these absent pairs are penalized:
\begin{equation}
 \hat{V}_2=\sum_{n_1,n_2=0}^{\infty}
 \sum_{M_{\rm rel}=M_\text{min,ele}^{n_1n_2}}^{M_\text{min,BCF}^{n_1n_2}-1}
 V_{M_{\rm rel}}^{n_1n_2}P_{n_1n_2}(M_{\rm rel})\non
\end{equation}
where $V_{M_{\rm rel}}^{n_1n_2}$ are positive numbers and 
$M_\text{min,ele}^{n_1n_2}=-n_1-n_2+\delta_{n_1n_2}$ is the minimum possible 
$M_{\rm rel}$ of two electrons from LLs $n_1$ and $n_2$ in the full Hilbert 
space of all fermionic many-body states. Using explicit expressions for 
$M_\text{min,ele}^{n_1n_2}$ and $M_\text{min,BCF}^{n_1n_2}$ here, one obtains
\begin{equation}
\hat{V}_2=\sum_{n_1=0}^{\infty}\sum_{n_2\neq n_1}
 V_{{M_{\rm rel}=-n_1-n_2}}^{n_1n_2}P_{n_1n_2}(-n_1-n_2).
 \label{eq:KSSJ-V2-DISK}
\end{equation}
This interaction obtains $\Psi_{1/2}^{(\alpha_1,\alpha_2)}$ as zero energy states. However, this contains interactions only between inter-LL pairs and thus, any wave function projected into the LLL also becomes a zero energy state. 
We need at least a three-body interaction to break this enormous degeneracy.

\subsection{Three body component $V_3$}
\label{sec:V3}
The three-body model component of the interaction is obtained in the same way as the two-body case. 
The minimum $M_{\rm rel}$ of three particles from the LLs $n_1,n_2,n_3$ of partitions $p_1,p_2,p_3$ in the state $\Phi_1^{\alpha_1}\Phi_1^{\alpha_2}$ is 
\begin{align}
-n_{123}+\sum_{i<j}^3\delta_{p_ip_j}\delta_{n_in_j},
\end{align}
where $n_{123}=\sum_{i=1}^3 n_i$. 
Acting $\hat{J}$ [Eq.~\eqref{eq:J(z_j,w_k)}] on $\Phi^{\alpha_1}_1\Phi_1^{\alpha_2}$ further increases this 
by $3+\delta_{p_1p_2}+\delta_{p_1p_3}+\delta_{p_2p_3}$ to 
\begin{align}
 &M^{n_1n_2n_3}_\text{min}(p_1,p_2,p_3)\non
 =&-n_{123}+\sum_{i<j}^3\delta_{p_ip_j}\delta_{n_in_j} +
 3+\delta_{p_1p_2}+\delta_{p_1p_3}+\delta_{p_2p_3}.
\end{align}
Thus, the minimum possible $M_\text{rel}$ in $\Psi_{1/2}^{(\alpha_1,\alpha_2)}$ is given by
\begin{align}
  M_\text{min,BCF}^{n_1n_2n_3}
 =&\min_{p_1,p_2,p_3}
 \left[
 M^{n_1n_2n_3}_\text{min}(p_1,p_2,p_3)
 \right]\non
 =&4-n_{123}+\delta_{n_1n_2}\delta_{n_2n_3}.
\end{align}
Figures~\ref{fig:Mmin_bcf}(c) and (d) show examples of triplets producing the $M_{\rm rel}=M_\text{min,BCF}^{n_1n_2n_3}$ channel.
The interaction for which $\Psi_{1/2}^{(\alpha_1,\alpha_2)}$ has zero 
interaction energy should eliminate all relative momenta $M_{\rm rel} < M_\text{min,BCF}^{n_1n_2n_3}$: 
\begin{align}
 \hat{V}_3
 &=\sum_{n_1,n_2,n_3=0}^{\infty}
 \sum_{M_{\rm rel}=M_\text{min,ele}^{n_1n_2n_3}}^{M_\text{min,BCF}^{n_1n_2n_3}-1}
 V_{M_{\rm rel}}^{n_1n_2n_3}P_{n_1n_2n_3}(M_{\rm rel}),
 \label{eq:KSSJ-V3-DISK}
\end{align}
where 
$M_\text{min,ele}^{n_1n_2n_3}=-n_{123}+\delta_{n_1n_2}+\delta_{n_1n_3}+\delta_{n_2n_3}$ is the smallest possible $M_\text{rel}$ for three electrons in the full Hilbert space [see Figs.~\ref{fig:Mmin_Full}(c)-(e)], and 
$P_{n_1n_2n_3}(M_{\rm rel})$ is the projection operator on states of three particles in LLs $n_1,n_2,n_3$ with a relative momentum of $M_{\rm rel}$.

Using the expressions for $M_\text{min,ele}^{n_1n_2n_3}$ and $M_\text{min,BCF}^{n_1n_2n_3}$ obtained above, we see that $\hat{V}_3$ interaction penalizes the following relative angular momentum channels:
\begin{enumerate}[(i)]
\setlength{\parskip}{0cm}
\setlength{\itemsep}{0cm}
\item For three particles in the same LL ($n_1=n_2=n_3$): $M_{\rm rel}=-n_{123}+3$ and $-n_{123}+4$.
\item For three particles in three different LLs ($n_1<n_2< n_3$):$-n_{123}\leq M_{\rm rel}\leq-n_{123}+3$.
\item When only two of the three are in the same LL ($n_1=n_2\neq n_3$): $-n_{123}+1\leq M_{\rm rel}\leq-n_{123}+3$.
\end{enumerate}
Figure~\ref{fig:pictorial} summarizes these interaction terms pictorially.

\subsection{Null space of the interaction}
\label{sec:null}
The wave function $\Psi_{1/2}^{(\alpha_1,\alpha_2)}$ in Eq.~\eqref{eq:KSSJ-psi}
is an eigenstate of $H$ in Eq.~\eqref{eq:KSSJ-H}, for the interaction constructed in the previous subsections, with zero interaction energy. 
Then, using $[\hat{Z}_k,\sum_j\hat{\bm{\pi}}_j^2/2m]=[\hat{W}_k,\sum_j\hat{\bm{\pi}}_j^2/2m]=0$, we obtain
\begin{align}
 \hat{H}\Psi^{(\alpha_1,\alpha_2)}_{1/2}
 =\left(E_{\alpha_1}+E_{\alpha_2}\right)\Psi^{(\alpha_1,\alpha_2)}_{1/2},
\end{align}
where $E_{\alpha_i}$ is the kinetic energy of $\Phi^{\alpha_i}_1$.

In particular, the Pfaffian ground state $\Psi^\text{Pf}_{1/2}$ and its QH
states $\Psi^\text{Pf,2n-qh}_{1/2}$ 
are reproduced by $\Psi^{(\alpha_1,\alpha_2)}_{1/2}$.
As a result, these states have zero interaction energy in our model. This can also be seen by noting that 
\begin{align}
\mathcal{P}_\text{LLL}(\hat{V}_2+\hat{V}_3)\mathcal{P}_\text{LLL} =\hat{V}_3^\text{Pf}.
 \label{eq:V2V3=Vpf}
\end{align}

As for the Abelian case in Ref.~\onlinecite{Anand21}, we conjecture that the states $\Psi^{(\alpha_1,\alpha_2)}_{1/2}$ in Eq.~\eqref{eq:KSSJ-psi} provide a complete basis for the zero interaction energy Hilbert space. 
An ``apparent" prediction for the number of degenerate states and their quantum numbers can be obtained by considering the ``parent" system $\Phi^{\alpha_1}_1(\{z_j\})\Phi^{\alpha_2}_1(\{w_j\})$, which would produce the correct answer provided 
that the linearly independent basis functions $\Phi^{\alpha_1}_1(\{z_j\})\Phi^{\alpha_2}_1(\{w_j\})$ produce linearly independent basis functions $\Psi^{\alpha_1,\alpha_2}_{1/2}$. However, we know that not to be the case for QHs, where the number of linearly independent basis functions is smaller than the apparent value~\cite{Sreejith11b}. Whether the same is true for the QPs is a priori not known, but would be crucial for establishing their non-Abelian braid statistics. This is one of the issues that we address by ED below.

\section{Numerical diagonalization}
\label{sec:num}
\subsection{Method}

Our numerical studies are performed in the standard spherical geometry~\cite{Haldane83}, where $N$ electrons move on the surface under a radial magnetic field. The total flux through this spherical surface is $2Q\phi_0$, where $\phi_0\equiv hc/e$ is the flux quantum and $2Q$ is an integer as required by the Dirac quantization condition. Rotational symmetry enables us to label the single-particle states by the orbital angular momentum $l$ and its $z$-component $m$. Their possible values are 
$l=|Q|,|Q|+1,|Q|+2,\ldots$ and $m=-l,-l+1,\ldots,l$, respectively. The $2l+1$ 
states with $l=|Q|+n$ form the $n$th LL. Many-body states are 
labeled by the total angular momentum $L$ as long as interactions are 
rotationally invariant.

By using the correspondence $L=2Q-M_\text{rel}$, where $M_\text{rel}$ is the 
relative angular momentum of the disk discussed above, the two-body interaction
in Eq.~\eqref{eq:KSSJ-V2-DISK} is transformed to 
\begin{align}
 \hat{V}_2
 &=\sum_{l_1=|Q|}^{\infty}\sum_{l_2\neq l_1}
 V_{L=l_1+l_2}^{l_1l_2}P_{l_1l_2}(L=l_1+l_2).
 \label{eq:KSSJ-V2-SPHERE}
\end{align}
In the same way but with the correspondence $L=3Q-M_\text{rel}$ for a triplet, 
Eq.~\eqref{eq:KSSJ-V3-DISK} is
transformed to
\begin{align}
 \hat{V}_3
 &=\sum_{l_1,l_2,l_3=|Q|}^{\infty}
 \sum_{L=L_\text{min,BCF}^{l_1l_2l_3}+1}^{L_\text{min,ele}^{l_1l_2l_3}}
 V_L^{l_1l_2l_3}P_{l_1l_2l_3}(L),
 \label{eq:KSSJ-V3-SPHERE}
\end{align}
where 
$L_\text{min,BCF}^{l_1l_2l_3}\equiv l_{123}-4-\delta_{l_1l_2}\delta_{l_2l_3}$,
$L_\text{min,ele}^{l_1l_2l_3}\equiv l_{123}-\delta_{l_1l_2}-\delta_{l_1l_3}-\delta_{l_2l_3}$, 
and we abbreviate $l_1+l_2+l_3=l_{123}$. 

\begin{figure*}[]
 \begin{center}
  \includegraphics[width=2\columnwidth]{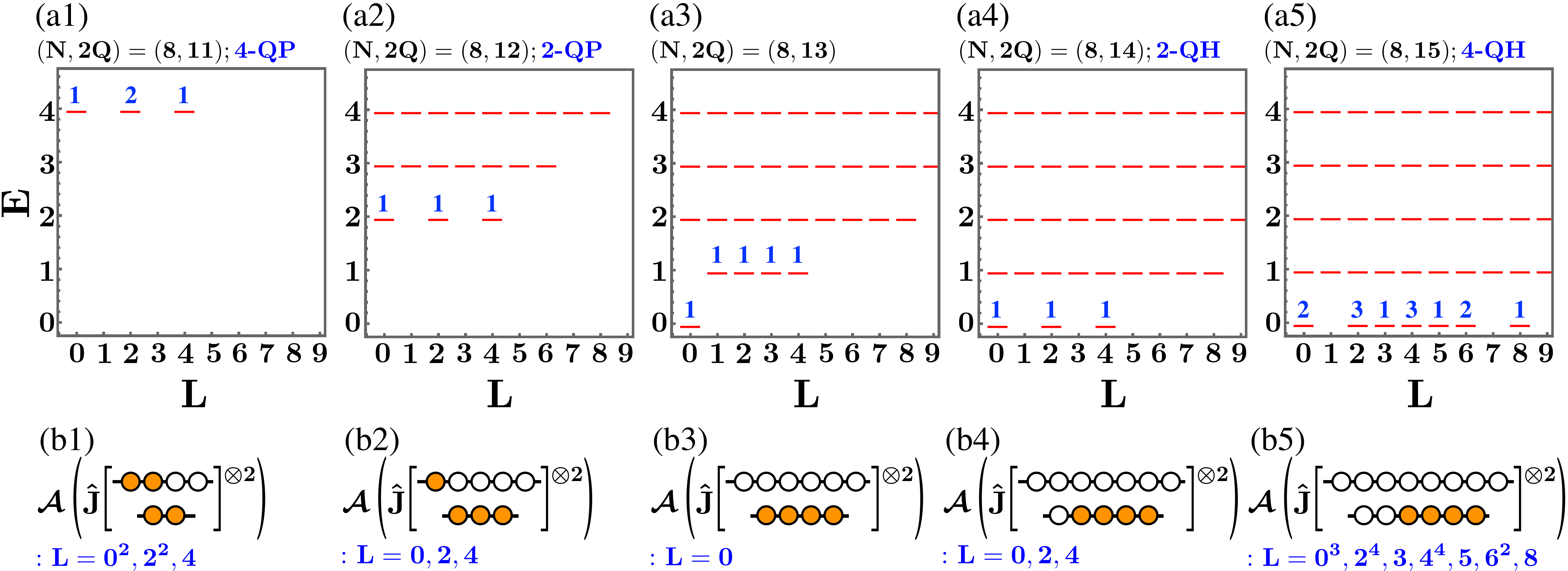}
 \end{center}
 \caption{
 (a1)-(a5) Energy spectra for $N=8$ electrons for $2Q=11,\ldots,15$ as a 
 function of the total angular momentum $L$. The number above a dash indicates 
 the degeneracy of that state. The energy $E$ is in units of $\hbar\omega_c$.
 (b1)-(b5) Schematic depiction of the lowest energy BCF state, with the  
 exponent ``$\otimes2$'' indicating two partitions. 
 The $L$ values of the lowest kinetic energy states as predicted by apparent BCF construction are also shown (which is what would be obtained if there were no linear dependencies for the basis functions). 
 }
 \label{fig:spectra1}
\end{figure*}
In second-quantized form, our total Hamiltonian [Eq.~\eqref{eq:KSSJ-H}] on a sphere is written as
$\mathcal{H}=\mathcal{H}_\text{kin}+\mathcal{V}_2+\mathcal{V}_3$, where
\begin{align}
 &\mathcal{H}_\text{kin}
 =\sum_{l=|Q|}^\infty\sum_{m=-l}^l
 (l-|Q|)\hbar\omega_c\hat{n}_{lm}
 \label{eq:Hkin}\\
 &\mathcal{V}_2
 =\frac{1}{2}\prod_{i=1}^{2}
 \left(
 \sum_{l_i,l'_i=|Q|}^\infty
 \sum_{m_i=-l_i}^{l_i}
 \sum_{m'_i=-l'_i}^{l'_i}
 \right)\times\non
 &\qquad\qquad\qquad\qquad
 \hat{c}^\dagger_{l_1m_1}\hat{c}^\dagger_{l_2m_2}
 V_{12;1'2'}\hat{c}_{l'_2m'_2}\hat{c}_{l'_1m'_1}\\
 &\mathcal{V}_3
 =\frac{1}{6}\prod_{i=1}^{3}
 \left(
 \sum_{l_i,l'_i=|Q|}^\infty
 \sum_{m_i=-l_i}^{l_i}
 \sum_{m'_i=-l'_i}^{l'_i}
 \right)\times\non
 &\qquad\qquad
 \hat{c}^\dagger_{l_1m_1}\hat{c}^\dagger_{l_2m_2}\hat{c}^\dagger_{l_3m_3}
 V_{123;1'2'3'}
 \hat{c}_{l'_3m'_3}\hat{c}_{l'_2m'_2}\hat{c}_{l'_1m'_1}.
\end{align}
In these expressions, $\hat{c}^\dagger_{lm}$ is the electron creation operator, $\hat{n}_{lm}=\hat{c}^\dagger_{lm}\hat{c}_{lm}$. 
In Eq.~\eqref{eq:Hkin}, we measure the kinetic energy relative to zero-point energy
$N\hbar\omega_c/2$ and also set, for simplicity, the separation between all successive LLs 
to be the same (for finite systems in the spherical geometry, the LL spacing depends slightly on the LL index). 
The symbols $V_{12;1'2'}$ and $V_{123;1'2'3'}$ are shorthands
for the matrix element, e.g., 
$V_{12;1'2'}\equiv\left(\bra{l_2,m_2}\otimes\bra{l_1,m_1}\right)\hat{V}_2
\left(\ket{l'_1,m'_1}\otimes\ket{l'_2,m'_2}\right)$.
These reduce to
\begin{widetext}
\begin{align}
 V_{12;1'2'}
 &=\delta_{l_1l_1'}\delta_{l_2l_2'}
 \sum_{l_1=|Q|}^{\infty}\sum_{l_2\neq l_1}
 \sum_{L_z=-L}^LV_L^{l_1l_2}
 C^{L,L_z}_{(l_1,m_1),(l_2,m_2)}C^{L,L_z}_{(l_1,m'_1),(l_2,m'_2)}
 \Bigr|_{L=l_1+l_2},\\
 V_{123;1'2'3'}
 &=\delta_{l_1l_1'}\delta_{l_2l_2'}\delta_{l_3l_3'}
 \sum_{l_1,l_2,l_3=|Q|}^{\infty}
 \sum_{L=L_\text{min,BCF}^{l_1l_2l_3}+1}^{L_\text{min}^{l_1l_2l_3}}
 \sum_{L_z=-L}^L\sum_{a}V_L^{l_1l_2l_3}
 C^{L,L_z,a}_{(l_1,m_1),(l_2,m_2),(l_3,m_3)}
 C^{L,L_z,a}_{(l_1,m'_1),(l_2,m'_2),(l_3,m'_3)},
\end{align}
\end{widetext}
where $C^{L,L_z}_{(l_1,m_1),(l_2,m_2)}$ and 
$C^{L,L_z,a}_{(l_1,m_1),(l_2,m_2),(l_3,m_3)}$ are the Clebsch Gordan
coefficients for two and three particles, and $a$ labels three-body degenerate 
states with same values of $(l_1,l_2,l_3,L,L_z)$; for example, the states with 
$L=l_1+l_2+l_3-3$ and $l_1=l_2\neq l_3$ are doubly degenerate apart from the 
$2L+1$ degeneracy. 

The Hamiltonian $\mathcal{H}$ conserves the total angular momentum $L$, its 
$z$-component $L_z$, and the particle number in each LL. Within the subspace 
specified by them, except for $L$, we diagonalize $H$ by using the Lanczos 
method. The ED is performed in the Hilbert space restricted 
in the lowest two LLs unless otherwise stated.

\subsection{Results}
Let us now discuss the spectra obtained from ED of 
the model described above. 
The $\nu=1/2$ Pfaffian state on a sphere occurs at $2Q=2N-3$.
In Figs.~\ref{fig:spectra1}(a1)-(a5), we plot the energy spectra in the
vicinity of $2Q=13$ with $N=8$. $L$ is the total angular momentum, and each dash 
represents a multiplet with $2L+1$ degenerate states; the degeneracy discussed below refers to the number of 
degenerate multiplets. We set $\hbar\omega_c$ to unity
and take all nonzero pseudopotentials to be sufficiently large so that states 
with nonzero interaction energies are pushed out of the figures. In 
Figs.~\ref{fig:spectra1}(a1) and (a2), the ground state with $2n$ QPs has
$E=2n\hbar\omega_c$, which matches the energy of the exact solution
$\Psi^\text{2n-qp}_{1/2}$ in Eq.~\eqref{eq:KSSJ-psi}. The number shown above a dash 
indicates the number of degenerate multiplets at that $L$ value. 
(We have confirmed that no new 
states appear at $E=2n\hbar\omega_c$ when we include the lowest four LLs in our 
ED study.) Figures~\ref{fig:spectra1}(b1)-(b5) show the BCF representations 
of these states, along with the apparent prediction for the $L$ values of the lowest QP and QH states. 
The degeneracy in the exact spectrum is not always 
in agreement with the apparent prediction from the BCF theory. 
While the apparent BCF 
prediction is correct for the ground state and the 2-QP / 2-QH states, it fails
for 4-QP / 4-QH states. The 4-QP state 
has $L=0,2^2,4$ in Fig.~\ref{fig:spectra1}(a1) (the superscript denotes the 
number of degenerate multiplets), to be compared to $L=0^2,2^2,4$ from the apparent BCF prediction. Similarly,
for the 4-QH state, ED produces $L=0^2,2^3,3,4^3,5,6^2,8$, whereas the apparent 
BCF counting gives $L=0^3,2^4,3,4^4,5,6^2,8$.
This implies nontrivial linear dependencies for the $\Psi^\text{2n-qp}_{1/2}$ basis functions.
(Note that the degeneracy of the $2n$-QH state 
 is greater than $2^{n-1}$, and also depends on $N$, because the positions of the QHs are not 
 fixed.)

This observation raises the question: What is the statistics of the QPs? Are the QPs Majoranas? If so, the degeneracies 
for QPs and QHs should be the same. Actually, the 2-QP and the 2-QH 
states [Figs.~\ref{fig:spectra1}(a2) and (a4)] have states with the same quantum numbers. However,
the 4-QP state in Fig.~\ref{fig:spectra1}(a1) exhibits a quite different 
structure from the 4-QH state in Fig.~\ref{fig:spectra1}(a5). 

\begin{figure}[t!!!]
 \begin{center}
  \includegraphics[width=\columnwidth]{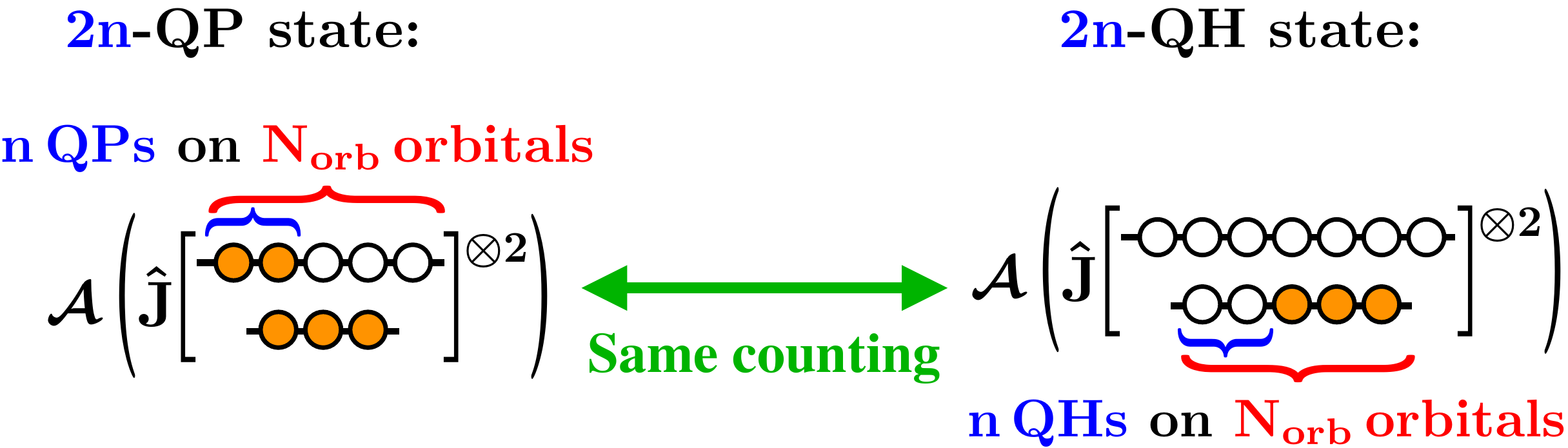}
 \end{center}
 \caption{Schematic view of a $2n$-QP state and the equivalent 
 $2n$-QH state. In this example, the parameters are set as $n=2$ and 
 $N_\text{orb}=5$. 
 }
 \label{fig:finding}
\end{figure}

\begin{figure}[t!!!]
 \begin{center}
  \includegraphics[width=\columnwidth]{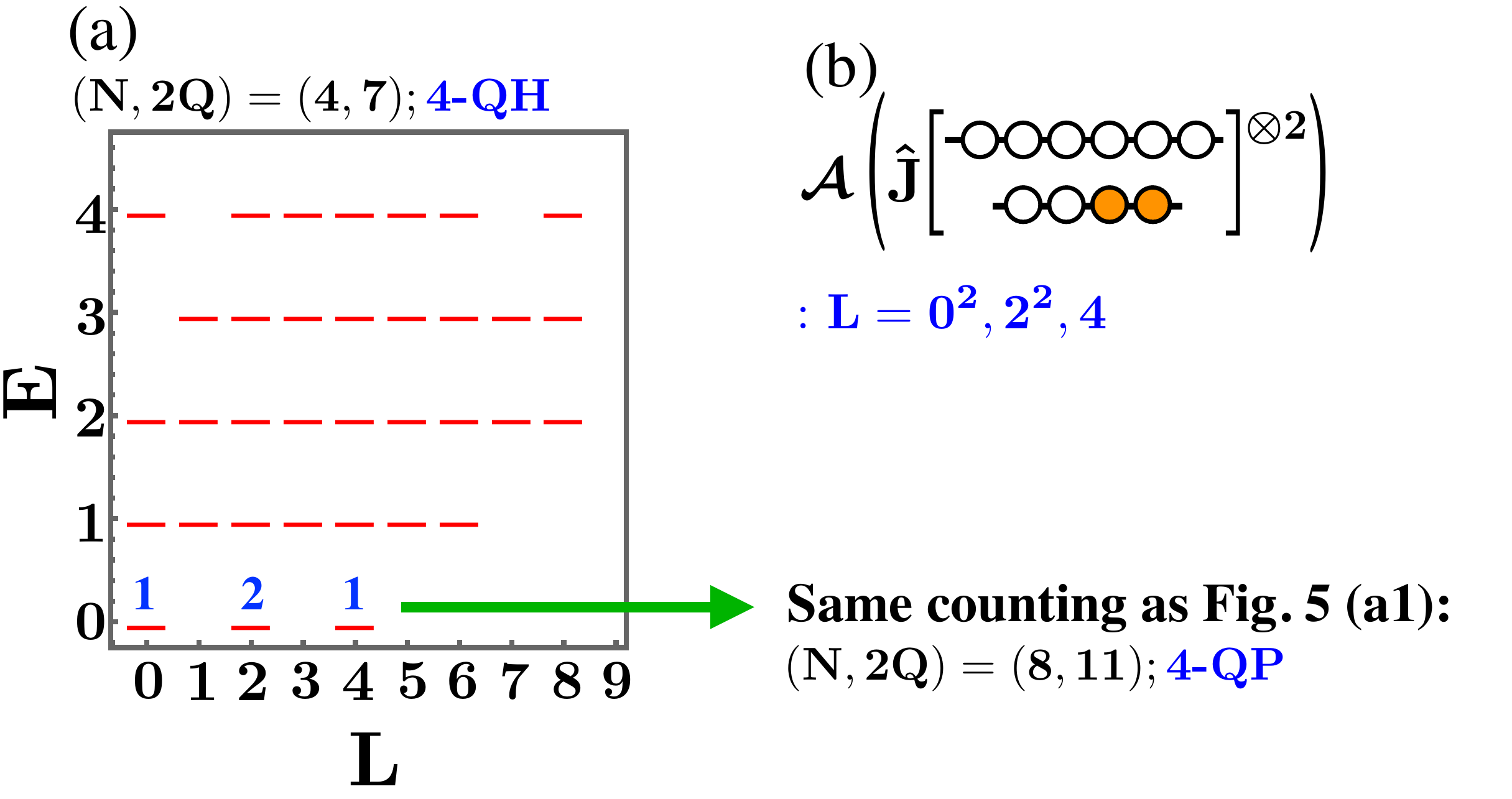}
 \end{center}
 \caption{
 (a) Spectrum with $(N,2Q)=(4,7)$. The exact counting is in agreement with that in
 Fig.~\ref{fig:spectra1}(a1). 
 (b) Schematic description of the lowest energy BCF state. Two QHs reside
 in the LLL in each partition, which should be compared to
 Fig.~\ref{fig:spectra1}(b1).
 }
 \label{fig:spectra2}
\end{figure}

The resolution to this apparent discrepancy is provided by the BCF model itself. A little thought shows that it is not appropriate to compare the 4-QP state in Fig.~\ref{fig:spectra1}(a1) with the 4-QH state in Fig.~\ref{fig:spectra1}(a5). That becomes clear by noting that in the BCF representations shown in Fig.~\ref{fig:spectra1}(b1) and Fig.~\ref{fig:spectra1}(b5), the two QHs in each partition are in a LL with $N_\text{orb}=6$, whereas the two QPs are in a LL with 
$N_\text{orb}=4$, where $N_\text{orb}$ is the total number of single particle orbitals. One must compare equivalent systems which are defined so that they have (i) the same number of QPs and QHs, and (ii) they reside in a LL with the same orbital degeneracy $N_\text{orb}$. (This correspondence was also noted by Rodriguez {\em et al.}~\cite{Rodriguez12b}.) Figure~\ref{fig:finding} gives an example of equivalent QP and QH systems. Specifically, 
the following two 
states are equivalent:
\begin{align}
 &\text{$2n$-QP state with $N=2N_\text{orb}+2n-4$}\non
 &\qquad\qquad\qquad\qquad\updownarrow\non
 &\text{$2n$-QH state with $N=2N_\text{orb}-2n$}.
 \label{eq:finding}
\end{align}
Our ED calculation shows (wherever we have been able to test) that equivalent QP and QH systems have the same counting for $2n=2$ and $2n=4$. For example, the system in 
Fig.~\ref{fig:spectra1}(a1) is equivalent through the above relations to a system with $N=4$. Figures~\ref{fig:spectra2}(a) and (b) show its spectrum and the schematic BCF 
depiction, clearly demonstrating the same counting as in Fig.~\ref{fig:spectra1}(a1).

\begin{table*}[]
 \caption{Number of the degenerate angular momentum $L$ multiplets in the ground state. Here $N$ is the number of electrons, and $2n$ is the number of QPs / QHs. 
The QP and QH systems in the same row are equivalent (related through Eq.~\eqref{eq:finding}). 
 The number in parentheses shows the ``apparent'' degeneracy predicted by the BCF approach whenever it is 
 different from the actual degeneracy. 
 $\dim H$ represents the dimension of the Hilbert 
 space with $L_z=0$ and $(N_\text{LLL},N_\text{SLL})=(N-2n,2n)$, where 
 $N_\text{LLL}$ and $N_\text{SLL}$ are the particle numbers in the lowest and
 second LLs, respectively.
 The result for the $2n$-QH state is the same as Table I in 
 Ref.~\onlinecite{Read96}. 
 }
 \label{tab:countingQPQH}
 \centering
 \begin{tabular*}{2.07\columnwidth}{@{\extracolsep{\fill}}ccccccccccccccccccccccc}
  \toprule
  \multicolumn{12}{c}{$2n$-QP state} & \multicolumn{11}{c}{$2n$-QH state} \\
  \cmidrule(lr){1-12}\cmidrule(lr){13-23}
  $N$ & $n$ & $L=0$ & 1 & 2 & 3 & 4 & 5 & 6 & 7 & 8 & $\dim H$ &
  $N$ & $n$ & $L=0$ & 1 & 2 & 3 & 4 & 5 & 6 & 7 & 8\\
  \cmidrule{1-12} \cmidrule{13-23} 
  \morecmidrules
  \cmidrule{1-12} \cmidrule{13-23} 
  4 & 1 & 1 &  & 1 &  &  &  &  &  &  & 26 & 4 & 1 & 1 &  & 1 &  &  &  & \\
  6 & 1 &  & 1 &  & 1 &  &  &  &  &  & 452 & 6 & 1 &  & 1 &  & 1 &  &  & \\
  8 & 1 & 1 &  & 1 &  & 1 &  &  &  &  & 7658 & 8 & 1 & 1 &  & 1 &  & 1 &  & \\
  10 & 1 &  & 1 &  & 1 &  & 1 &  &  &  & 126510 & 10 & 1 &  & 1 &  & 1 &  & 1 & \\
  12 & 1 & 1 & & 1 & & 1 & & 1 & &  & 2069194  & 12 & 1 & 1 &  & 1 &  & 1 &  & 1 \\
  14 & 1 &  & 1 &  & 1 &  & 1 &  & 1 &  &  33630328 & 14 & 1 &  & 1 &  & 1 &  & 1 &  & 1\\
  6 & 2 & 1 &  & 1 &  &  &  &  &  &  & 410 & 2 & 2 & 1 &  & 1 &  &  &  &  \\
  8 & 2 & 1(2) &  & 2 &  & 1 &  &  &  &  & 21007 & 4 & 2 & 1(2) &  & 2 &  & 1 &  &  \\
  10 & 2 & 2 &  & 2(3) & 1 & 2 &  & 1 &  &  & 728380 & 6 & 2 & 2 &  & 2(3) & 1 & 2 &  & 1 \\
  12 & 2 & 2(3) & & 3(4) & 1 & 3(4) & 1 & 2 &  & 1 & 20691552 & 8 & 2 & 2(3) &  & 3(4) & 1 & 3(4) & 1 & 2 & & 1 \\
  10 & 3 &  & 1 &  & 1 &  &  &  &  &  & 527102 & 2 & 3 &  & 1 &  & 1 &  &  &  \\
  12 & 3 & 2 &  & 2(3) & 1 & 2 &  & 1 &  &  & 33699452 & 4 & 3 & 1(2) &  & 2(3) & 1 & 2 &  & 1 \\
  \midrule
  \bottomrule
 \end{tabular*}
\end{table*}
We have performed ED for many systems and summarize the
results in Table~\ref{tab:countingQPQH}. We can compare the spectra of equivalent QP and QH systems. They are in complete agreement for 2-QP and 4-QP states. For the 2-QP states, we have the same
counting as apparently predicted by the BCF construction. In other words, we obtain same states as our exact solutions $\Psi^\text{2-qp}_{1/2}$ in Eq.~\eqref{eq:KSSJ-psi} 
as we conjectured. For 4-QP states, we obtain a counting
smaller than the apparent BCF prediction. This implies a linear dependence of 
$\Psi^\text{4-qp}_{1/2}$ basis functions as mentioned above, which, 
interestingly, results in the same counting of states as that for the QHs. For 6-QP states, 
we again find linear dependencies, producing a smaller set of states than the apparent BCF 
prediction.
In this case, however, the counting of states is not identical to that of the 
equivalent 6-QH system. 
This behavior is in agreement with that found in Ref.~\cite{Rodriguez12b}.

\begin{figure}[]
\includegraphics[width=\columnwidth]{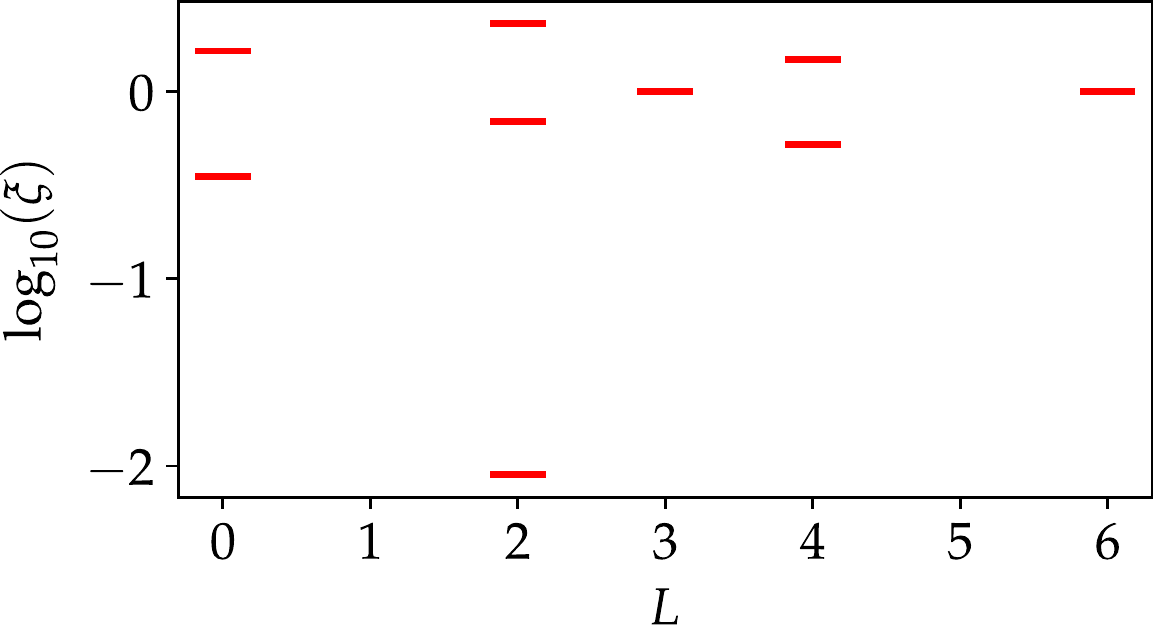}
\caption{Eigenvalues $\xi$ of the overlap matrix of 6-QP BCF states (Eq.~\ref{eq:BCFwf}) projected into the LLL by the Jain-Kamilla method for different total angular momentum channels $L$ (all in the $L_z=0$ sector). While the apparent counting at $L=2$ is 3,
one of the states is approximately linearly dependent on the others as indicated by a low lying eigenvalue here. 
(The linear dependence is rendered exact if exact projection is considered.)\label{fig:BCF_SVD}}
\end{figure}

To give an example, 
the 6-QP state with $N=12$ in Table~\ref{tab:countingQPQH} has 
$L=0^2,2^2,3,4^2,6$. We compare it with the number of approximately linear 
indepdnent states in a set of the BCF trial wave functions
$\Psi^\text{BCF,6-qp}_{1/2}$ in Eq.~\eqref{eq:BCFwf} using
the Jain-Kamilla method. This number is 
inferred from the significant singular values of their overlap matrix.
As shown in Fig.~\ref{fig:BCF_SVD}, the two countings match exactly at 
each angular momentum.

Independent of the comparison with the equivalent QH systems, the primary conclusion is that {\it the counting of QP states, of our model Hamiltonian is always consistent with that predicted by the BCF model with exact LLL projection for all cases we have considered.} The conclusions made by Rodriguez {\em et al.}~\cite{Rodriguez12b} on the basis of the BCF model also apply to our exact Hamiltonian. In particular, the edge spectra, which correspond to large angular momentum states, are consistent with the MR prediction.  

The counting of the QP states of the model Hamiltonian considered here is consistent with a Majorana mode associated with each QP. So far, we have demonstrated non-Abelian statistics of QPs. Let 
us now discuss neutral excitations in terms of our exact solutions.
\begin{table}[t]
 \caption{Number of the degenerate angular momentum $L$ multiplets for an ordinary exciton. The total flux is $2Q=2N-3$. 
 $\dim H$ represents the dimension of 
 the Hilbert space with $L_z=0$ and $(N_\text{LLL},N_\text{SLL})=(N-1,1)$.
 }
 \label{tab:countingEx}
 \centering
 \begin{tabular*}{\columnwidth}{@{\extracolsep{\fill}}ccccccccc}
  \toprule
  $N$ & $L=1$ & 2 & 3 & 4 & 5 & 6 & 7 & $\dim H$\\
  \midrule\midrule
  4 & 1 & 1 &  &  &  &  &  & 18\\
  6 & 1 & 1 & 1 &  &  &  & & 196 \\
  8 & 1 & 1 & 1 & 1 &  &  &  & 2342\\
  10 & 1 & 1 & 1 & 1 & 1 &  &  & 29828\\
  12 & 1 & 1 & 1 & 1 & 1 & 1 &  & 396126\\
  14 & 1 & 1 & 1 & 1 & 1 & 1 & 1 & 5415354\\
  \midrule
  \bottomrule
 \end{tabular*}
\end{table}

The ordinary exciton is described in the BCF description as a QP-QH pair in one
partition, which implies that this state occurs in our 
Hamiltonian at $2Q=2N-3$ with energy $E=\hbar\omega_c$. 
In Fig.~\ref{fig:spectra1}(a3) with $N=8$, the states with $E=\hbar\omega_c$ 
have $L=1,2,3,4$. Table~\ref{tab:countingEx} summarizes other results for 
various systems with $4\leq N\leq14$. In all cases, we find ordinary
excitons at $L=1,2,\ldots,N/2$. This counting is consistent with the BCF 
prediction, which confirms that $\Psi^\text{1-exciton}_{1/2}$ in 
Eq.~\eqref{eq:KSSJ-psi} at each $L$ is a unique solution  of this model at energy $\hbar\omega_c$, as we 
conjectured.

\begin{table}[t]
 \caption{Number of the degenerate angular momentum $L$ multiplets for a topological exciton. The total flux is $2Q=2N-3$. 
 $\dim H$ represents the dimension of 
 the Hilbert space with $L_z=0$ and $(N_\text{LLL},N_\text{SLL})=(N-1,1)$}.
 \label{tab:countingTEx}
 \centering
 \begin{tabular*}{\columnwidth}{@{\extracolsep{\fill}}cccccccccc}
  \toprule
  $N$ & $L=1/2$ & 3/2 & 5/2 & 7/2 & 9/2 & 11/2 & 13/2 & 15/2 & $\dim H$\\
  \midrule\midrule
  3 & 1 & 1 &  &  &  &  &  &  & 6\\
  5 & 1 & 1 & 1 &  &  &  &  &  & 59\\
  7 & 1 & 1 & 1 & 1 &  &  &  &  & 670\\
  9 & 1 & 1 & 1 & 1 & 1 &  &  &  &  8298\\
  11 & 1 & 1 & 1 & 1 & 1 & 1 &  &  & 108182\\  
  13 & 1 & 1 & 1 & 1 & 1 & 1 & 1 &  & 1459692\\  
  15 & 1 & 1 & 1 & 1 & 1 & 1 & 1 & 1 & 20185883\\  
  \midrule
  \bottomrule
 \end{tabular*}
\end{table}
The topological exciton is also a QP-QH pair like a ordinary one, but the QP 
and QH reside in a different partition. This state occurs in our Hamiltonian 
with odd $N$ and $2Q=2N-3$ with $E=\hbar\omega_c$ as the lowest energy
state. Table~\ref{tab:countingTEx} summarizes the results for systems
with $3\leq N\leq15$. In all cases, we find topological excitons at 
$L=1/2,3/2,\ldots,N/2$. As in the case of ordinary excitons, the counting is 
consistent with the BCF prediction, i.e., our exact solution at each $L$ is a 
unique solution of this model, as we conjectured.

\section{Quasiparticle charge}
\label{sec:fcharge}
We next discuss the local charge of a QP in our model. Conceptually, one can show that the QPs (QHs) 
have a local charge of magnitude $1/4$ of an electron charge 
straightforwardly by noting that 8 QPs (QHs) are generated in the BCF 
state when one adds (removes) two electrons~\cite{Sreejith11b}.
We now explicitly evaluate the QP charge by using our exact wave functions. 

\begin{figure}[t]
 \begin{center}
  \includegraphics[width=\columnwidth]{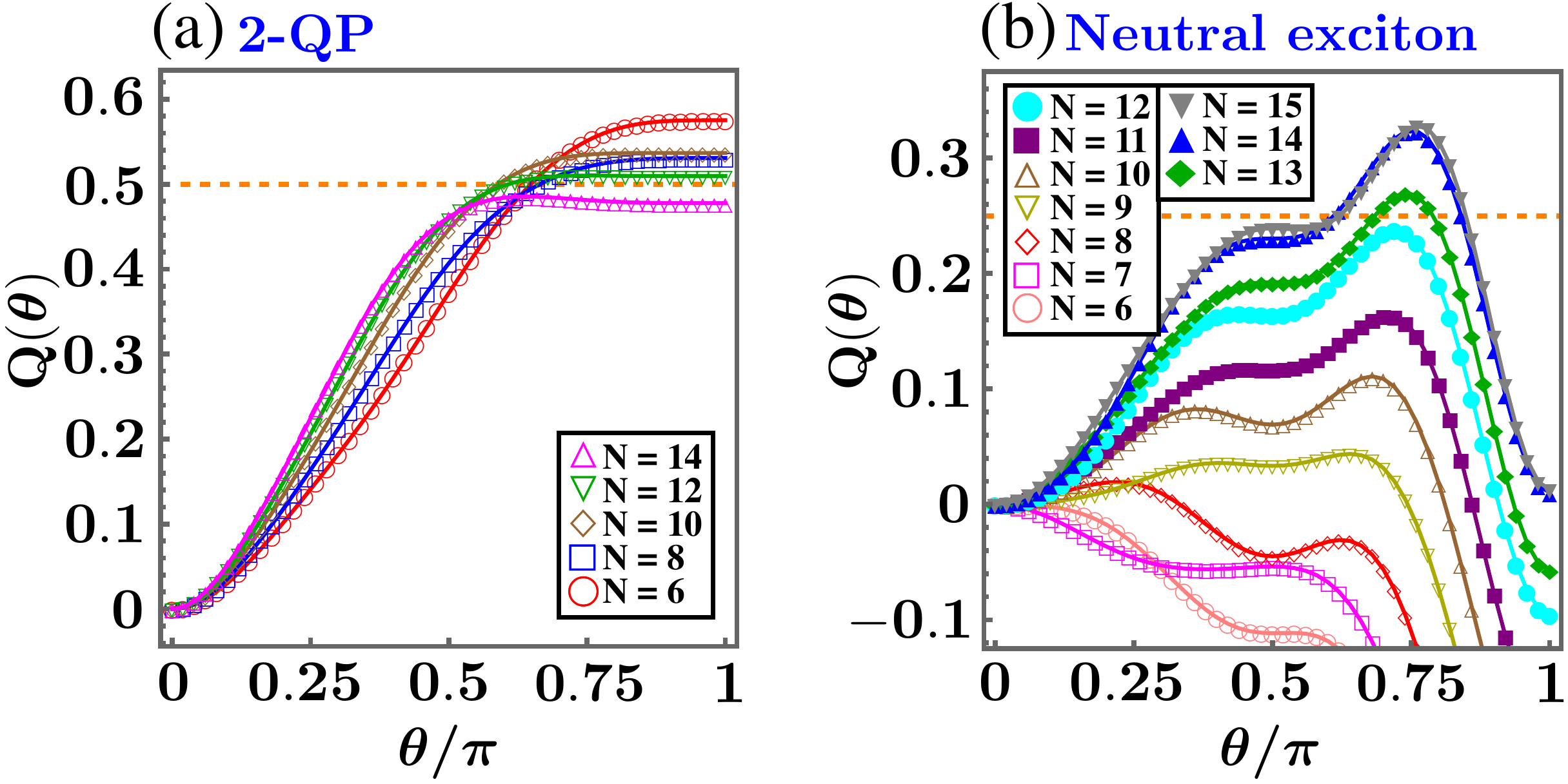}
 \end{center}
 \caption{
 Cumulative charge $Q(\theta)$ for (a) two QPs and (b) an ordinary exciton ($N$ even)
 or a topological exciton ($N$ odd). The $x$-axis is the polar angle 
 $\theta$. The horizontal dashed orange lines indicate (a) $Q(\theta)=1/2$ and (b) 
 $Q(\theta)=1/4$.
 }
 \label{fig:Q}
\end{figure}

We first consider the system with $2Q=2N-4$, where 2-QP states occur at 
$L=N/2,N/2-2,\ldots$ as shown in Table~\ref{tab:countingQPQH}. In the BCF 
description, the two QPs with $L=L_z=N/2$ are localized at the north pole. 
Their existence results in excess charge in the vicinity of the north pole, which can be 
calculated as the saturation value of~\cite{Anand21}
\begin{align}
 Q(\theta)
 =2\pi\int_0^\theta
 d\theta'\sin\theta'
 \left[
 \rho(\theta')-\rho_0
 \right],
 \label{eq:Q}
\end{align}
where $\rho(\theta)$ is the local charge density at the polar angle $\theta$
and $\rho_0\equiv\rho(\pi)$. (The charge density is independent of the
azimuthal angle $\phi$ due to rotational symmetry around the $z$-axis.) 
Figure~\ref{fig:Q}(a)
shows $Q(\theta)$ for various systems. The value of $Q(\theta)$ 
approaches 0.5 at $\theta/\pi\approx0.7$ and then saturates. This is consistent
with the fact that two QPs exist at the north pole and each of them has a 
charge of 1/4. As the particle number $N$ increases, $Q(\theta)$
begins to saturate faster.
The slight deviation of charge from 0.5 is a finite size effect.

Let us next consider states with a neutral exciton. As shown in 
Tables~\ref{tab:countingEx} and \ref{tab:countingTEx}, these states occur at
$L=N/2,N/2-1,\ldots$. When $L=L_z=N/2$, an exciton is described as a pair of 
a QP at the north pole and a QH at the south pole. Approximating $\rho_0$ by $\rho(\pi/2)$
in Eq.~\eqref{eq:Q}, we plot $Q(\theta)$ for various systems in 
Fig.~\ref{fig:Q}(b). In the thermodynamic limit,
$Q(\theta)$ would be flat with the value 0.25 in the middle region of 
$\theta$ and 
$Q(0)=Q(\pi)=0$. 
Although such a flat region is not clearly seen in our results, 
implying that the QH and QP sizes are larger than half the perimeter of the sphere,
the charge at $\theta/\pi\sim0.5$ is beginning to saturate at 0.25
in the large systems.

\section{Adiabatic continuity}
\label{sec:adiabatic}

We now ask if our exact solutions with QPs or neutral excitons are 
adiabatically connected to states of 
$\hat{V}_3^\text{Pf}$ in the LLL. While numerical solutions of 
the LLL $\hat{V}_3^\text{Pf}$ problem do not show a clear separation of 
energy scale for QPs, one obtains clear branches of states with an ordinary or
a topological exciton~\cite{Sreejith11b,Sreejith11}. We investigate 
adiabatic continuity for these neutral excited states.

One can continuously deform our model Hamiltonian into the LLL 
$\hat{V}_3^\text{Pf}$ interaction simply by increasing the cyclotron energy 
$\hbar\omega_c$ to a limit where it is much larger than the interaction energy. This is sufficient for the demonstration of the adiabatic continuity of the MR Pfaffian ground state and its QH states, which remain at zero energy during this process. The situation is more complicated for states with non-zero kinetic energies, however. We consider here the states with an ordinary or a topological exciton. As we increase $\hbar\omega_c$, these states 
continue to remain exact solutions, while 
simply floating up in energy and eventually crossing 
states where all particles occupy the LLL. 
To convert these level crossings into anti-crossings, we introduce a two-body interaction that breaks the conservation of 
particle number in each LL:
\begin{align}
 &\hat{V}'=\sum_{l_1,l_2,l'_1,l'_2=|Q|}^\infty
 V'^{l_1l_2l'_1l'_2}_{L_\text{max}} P_{l_1l_2,l'_1l'_2}(L_\text{max}),
\end{align}
where $P_{l_1l_2,l'_1l'_2}(L)
=\sum_{L_z=-L}^{L}\ket{l_1,l_2,L,L_z}\bra{l'_1,l'_2,L,L_z}$, 
$L_\text{max}=\min\{l_1+l_2-\delta_{l_1l_2},l'_1+l'_2-\delta_{l'_1l'_2}\}$, and
we set $V'^{|Q||Q||Q||Q|}_{L_\text{max}}=0$ to remove two-body interactions within the
LLL. We denote the second quantization representation of this interaction by $\mathcal{V}'$. 

\begin{figure}[t]
 \begin{center}
  \includegraphics[width=\columnwidth]{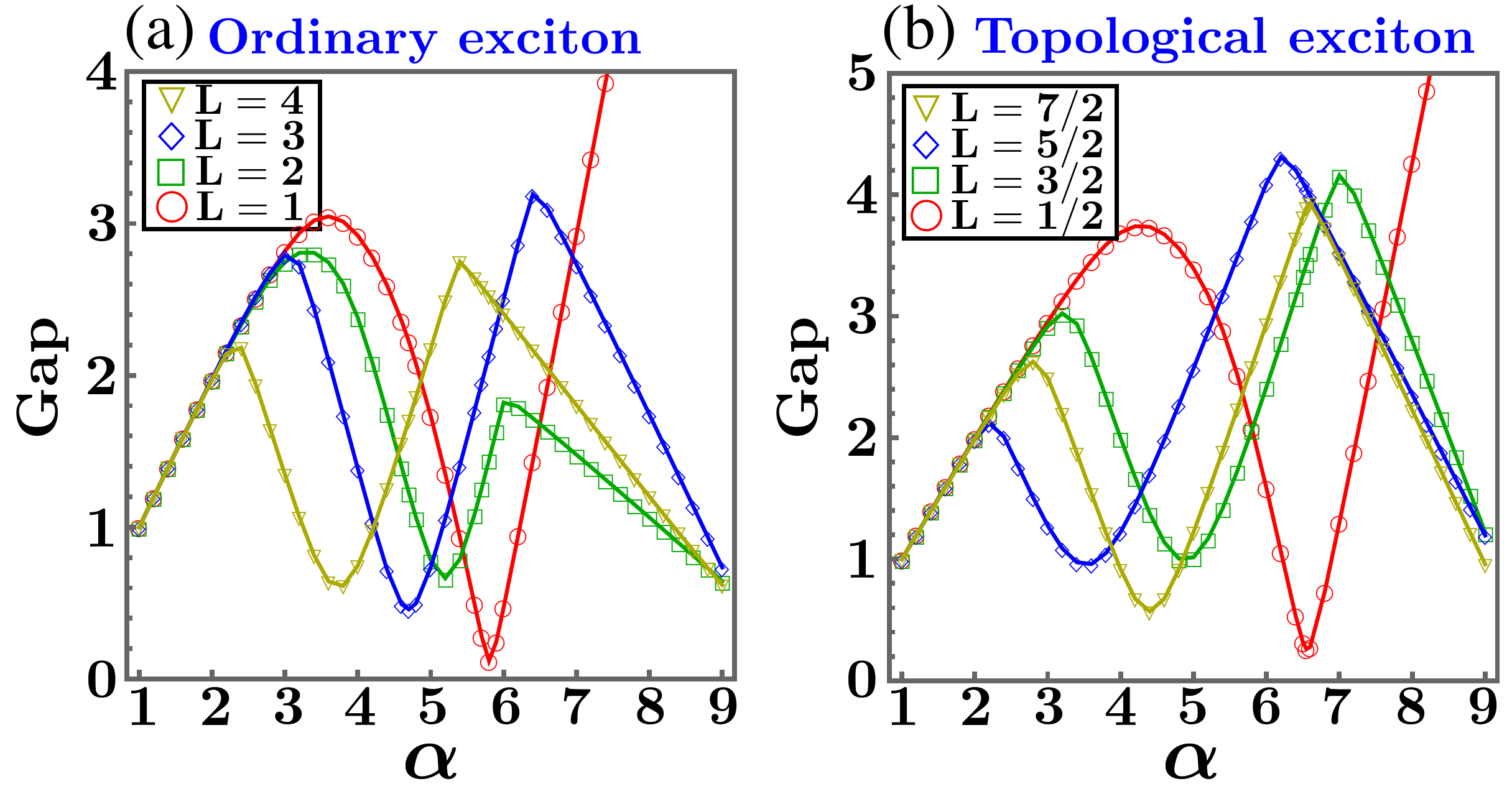}
 \end{center}
 \caption{
 Energy gap separating the two lowest energy states in various $L$ sectors as a function of the tuning parameter $\alpha$ for (a) an ordinary neutral 
 exciton for an $N=8$ particle system, and (b) a topological exciton for an $N=7$ particle system.
 }
 \label{fig:adiabatic}
\end{figure}
In Fig.~\ref{fig:adiabatic}, we plot the evolution of the gap for a neutral 
and a topological exciton at all relevant $L$ values for the  
Hamiltonian:
\begin{align}
 \mathcal{H}(\alpha)=\alpha\mathcal{H}_\text{kin}
 +(10-\alpha)\left(\mathcal{V}_2+\mathcal{V}_3\right)
 +\frac{\alpha-1}{8}\mathcal{V}',
\end{align}
where $\hbar\omega_c$ and all nonzero pseudopotentials are set to unity. 
$\mathcal{H}(1)
=\mathcal{H}_\text{kin}+9\left(\mathcal{V}_2+\mathcal{V}_3\right)$ has 
strong interactions so that our exact solutions become the ground states, whereas 
$\mathcal{H}(9)=9\mathcal{H}_\text{kin}+\mathcal{V}_2+\mathcal{V}_3+\mathcal{V}'$ 
has large kinetic energy and is thus effectively equivalent to
 $\hat{V}_3^\text{Pf}$. The 
particle number in each LL is not conserved in the intermediate region. 
The results in Figs.~\ref{fig:adiabatic}(a) and (b) demonstrate adiabatic
continuity for the ordinary and the topological exciton, respectively. The only exceptions
are the states with the smallest $L$ [i.e., $L=1$ in Fig.~\ref{fig:adiabatic}(a) 
and $L=1/2$ in Fig.~\ref{fig:adiabatic}(b)], where this gap appears to close, which  
is consistent with the absence of excitons at these angular momenta for the LLL 
$\hat{V}_3^\text{Pf}$ interaction; the BCF wave functions at these angular 
momenta are annihilated upon LLL projection, as noted in Refs.~\cite{Sreejith11b,Sreejith11}.

\section{Concluding remarks}

In summary, we have constructed an exactly solvable
model that produces QPs of the Pfaffian state as exact
eigenstates. The counting of states in the QP Hilbert
space has been found to be in 
agreement with that predicted from the BCF wave functions~\cite{Rodriguez12b}.
This provides a compelling demonstration, for a model interaction, that the QPs of the $\nu=1/2$ state are associated with Majorana modes. We have also numerically confirmed adiabatic continuity for states with a neutral exciton, as we deform our model Hamiltonian
into the LLL $\hat{V}^{\rm Pf}$ interaction. Whether the adiabatic 
continuity of the QHs and QPs of our model (or of the $\hat{V}^{\rm Pf}$
interaction) extends to the QHs and QPs of the Coulomb interaction remains an important open problem. Unfortunately, small system studies are not likely to be sufficient to address this question as the sizes of the QPs and QHs are comparable to the system sizes currently accessible.

It is tempting to ask if the degeneracies for the QP states can be calculated analytically following the work of Read and Rezayi for the QH states~\cite{Read96}. We have not succeeded in that goal. It may be noted that the QP wave functions are more complicated than the QH wave functions, as they involve more than one LL and a Jastrow operator rather than a Jastrow factor. Our QP wave functions also cannot be expressed in a Pfaffian form, which played a crucial role in Ref.~\cite{Read96}. The last point also underscores the usefulness of the BCF form in our construction of the exact solutions for the excitations.

Our work lends itself to generalization in many directions, which we mention here, but whose 
detailed study is left for the future. 
Although we have focused on $\nu=1/2$ and its vicinity, our formulation is valid for all BCF states 
and thus applies to 
$\nu=2\nu^*/[(2p+1)\nu^*+1]$~\cite{Sreejith11b}, where $\nu^*$ is arbitrary. In particular, it will produce 
non-Abelian states at $\nu=2n /[(2p+1)n+1]$, which, for $p=1$, correspond to $\nu=1/2$, 4/7, 3/5, $\cdots$. 
The issue of the non-Abelian statistics of the QHs / QPs of these states can be in principle studied by 
our method. 
An exact model for the QPs of the 
$\mathbb{Z}_k$ Read-Rezayi states~\cite{Read99} can also be constructed 
using the $k$-partite CF description including up to $k+1$-body interactions (e.g., see \cite{Sreejith13}).
Abelian and non-abelian states can also be constructed from the parton theory of the FQHE~\cite{Jain89b,Wen91}. For 
many of these states, exact Hamiltonians have been constructed~\cite{Jain90,Wu17,Chen17,Bandyopadhyay20,Ahari22}. It 
remains an open problem whether our approach can suggest an alternative construction for parent Hamiltonians of these states.

\begin{acknowledgments}
K.K. thanks JSPS for support from Overseas Research Fellowship.
A.S. and J.K.J acknowledge financial support from the U.S. National Science Foundation under grant no. DMR-2037990. G.J.S. thanks Abhishek Anand for useful inputs and DST-SERB (India) grant ECR/2018/001781 for financial support.
We acknowledge Advanced CyberInfrastructure computational resources provided by The Institute for CyberScience at The Pennsylvania State University. 

\end{acknowledgments}

\bibliography{biblio_fqhe}
\bibliographystyle{apsrev}

\end{document}